\documentclass[aps,prd,twocolumn,nofootinbib,showpacs,longbibliography]{revtex4-2}
\usepackage{bm,amsmath,amssymb,graphicx}
\usepackage[colorlinks,linkcolor=blue,urlcolor=blue]{hyperref}
\newcommand{\Ubar}[1]{\underline{#1}}
\newcommand{\dUbar}[1]{\underline{#1} \hspace{.6pt} \underline{#1}}
\newcommand{\Obar}[1]{\overline{#1}}
\newcommand{\dObar}[1]{\overline{#1} \hspace{.6pt} \overline{#1}}

\begin{document}

\title{From freely falling frames to the Lorentz gauge-symmetry group and a Hamiltonian composite theory of gravitation}

\author{Hans Christian \"Ottinger}
\email[]{hco@mat.ethz.ch}
\homepage[]{www.polyphys.mat.ethz.ch}
\affiliation{ETH Z\"urich, Quantum Center and Department of Materials, HCP F 43.1, CH-8093 Z\"urich, Switzerland}

\date{\today}

\begin{abstract}
The concept of freely falling frames suggests that gravity exhibits a local Lorentz gauge symmetry and requires a background Minkowski reference frame. The gauge vector fields of a Yang-Mills-type theory can be constructed from the transformations to these local freely falling frames, thereby leading to a composite theory of gravity. We propose coordinate conditions under which an exact black-hole solution can be obtained. Our analysis of planar gravitational waves reveals that, despite the large symmetry group, composite gravity possesses only four physical degrees of freedom. We elaborate a Hamiltonian formulation of composite gravity, derive the full set of constraints for the nonlinear theory, and outline the pathway toward its quantization.
\end{abstract}

\maketitle

\section{Introduction}
On Thursday, January 27, 1921, Albert Einstein gave a ceremonial address on the topic of ``Geometry and Experience'' to the Prussian Academy of Sciences in Berlin. In the following year, an expanded version of his German lecture \cite{EinsteinD} was translated into English \cite{EinsteinE}.
Einstein stated: ``Geometry (G) predicates nothing about the relations of real things, but only geometry together with the purport (P) of physical laws can do so. Using symbols, we may say that only the sum of (G) + (P) is subject to the control of experience. Thus (G) may be chosen arbitrarily, and also parts of (P); all these laws are conventions. All that is necessary to avoid contradictions is to choose the remainder of (P) so that (G) and the whole of (P) are together in accord with experience. Envisaged in this way, axiomatic geometry and the part of natural law which has been given a conventional status appear as epistemologically equivalent.''
% "Die Geometrie (G) sagt nichts über das Verhalten der wirklichen Dinge aus, sondern nur die Geometrie zusammen mit dem Inbegriff (P) der physikalischen Gesetze. Symbolisch können wir sagen, daß nur die Summe (G) + (P) der Kontrolle der Erfahrung unterliegt. Es kann also (G) willkürlich gewählt werden, ebenso Teile von (P); alle diese Gesetze sind Konventionen. Es ist zur Vermeidung von Widersprüchen nur nötig, den Rest von (P) so zu wählen, daß (G) und das totale (P) zusammen den Erfahrungen gerecht werden. Bei dieser Auffassung erscheinen die axiomatische Geometrie und der zu Konventionen erhobene Teil der Naturgesetze als erkenntnistheoretisch gleichwertig."

Einstein's endorsement of geometric conventionalism clearly contradicts the widespread belief that pseudo-Riemannian geometry of spacetime is the sacrosanct essence of gravity. This extraordinary quote is not mentioned in the influential textbooks on gravitation of the 1970s. For example, Weinberg \cite{Weinberg} does not cite Einstein's lecture and Misner, Thorne and Wheeler \cite{MisnerThorneWheeler} merely offer the following catchy quote from it: ``As far as the laws of mathematics refer to reality, they are not certain; and as far as they are certain, they do not refer to reality.''

Einstein's equivalence principle, which implies the observed equivalence of gravitational and inertial mass, is based on the physical equivalence of a gravitational field and a corresponding acceleration of the reference frame. This idea is often expressed along the following lines: No local experiment can distinguish a freely falling non-rotating frame in the presence of a gravitational field from a uniformly moving frame in the absence of a gravitational field. Or: Life in a laboratory falling freely in an external gravitational field is equivalent to life in the absence of external gravity, where variations of the gravitational field on the length scale of the size of the laboratory are assumed to be negligible.

Before and after the completion of general relativity, Einstein's papers \cite{Einstein11,Einstein16} provide arguments in favor of the existence of a background Minkowski space. In \cite{Einstein11}, he states the physical equivalence of a frame at rest in a homogeneous gravitational field and a uniformly accelerated frame in a space free of gravitational fields; in the absence of gravitational fields, however, special relativity is applicable and provides a background Minkowski space for reference. In \cite{Einstein16}, Einstein assumes that massive particles sufficiently far away from the sources of a gravitational field move uniformly; the asymptotic space-time may be regarded as a background Minkowski space. Note that free fall and accelerated motion must always be defined with respect to a background frame.

As the idea of freely falling local frames is at the heart of gravity, it is natural to introduce the matrices ${b^\kappa}_\mu$ describing the transformation from a background Minkowski frame to freely falling local frames, where the indices $\kappa$ and $\mu$ are associated with the freely falling and background frames, respectively. In the first place, freely falling frames are defined along the trajectory of a freely falling observer. By varying the initial conditions, one can cover the entire space. A smooth field of matrices ${b^\kappa}_\mu$ can be introduced by the same arguments as a metric in general relativity. Actually the decomposition
\begin{equation}\label{localinertialdecomp}
   g_{\mu\nu} = \eta_{\kappa\lambda} \, {b^\kappa}_\mu {b^\lambda}_\nu ,
\end{equation}
which may be interpreted as the transformation of the Minkowski metric $\eta_{\kappa\lambda} = \eta^{\kappa\lambda}$ with signature $(-,+,+,+)$ of the freely falling frame to a pseudo metric $g_{\mu\nu}$ in the background space, offers a simple possibility to construct the field ${b^\kappa}_\mu$ from the metric of general relativity. The fields ${b^\kappa}_\mu$ are known as tetrad, frame, or vierbein fields.

Because they forthrightly capture the notion of freely falling frames, we take the tetrad variables to be the fundamental variables in a theory of gravitation. The derived ``metric'' (\ref{localinertialdecomp}) defines a relation between co- and contra-variant vectors. In minimal terms, it may be interpreted as an anisotropic relation between the velocity and energy-momentum four-vectors in the background space, arising from an isotropic relation in the freely falling frames \cite{hco231}.

A local Lorentz transformation maps one freely falling frame to an equivalent one. Hence, a theory of gravity based on tetrad variables possesses a local gauge symmetry. A gauge transformation is given by
\begin{equation}\label{gaugetransformb}
   {b^\kappa}_\mu \rightarrow {\Lambda^\kappa}_\lambda \, {b^\lambda}_\mu \,,
\end{equation}
where ${\Lambda^\kappa}_\lambda$ represents a local Lorentz transformation. The invariance of $g_{\mu\nu}$ follows from the invariance of the Minkowski metric under Lorentz transformations,
\begin{equation}\label{Minkowskiinvariance}
    \eta_{\kappa'\lambda'} \, {\Lambda^{\kappa'}}_\kappa \, {\Lambda^{\lambda'}}_\lambda
    = \eta_{\kappa\lambda} \,.
\end{equation}

The local Lorentz symmetry must be distinguished from the global Lorentz symmetry of the background Minkowski space. As there is no preferential origin of this Minkowski space, the global symmetry is actually with respect to the Poincare group, which includes space-time translations. For the local symmetry at some point in space-time, it is natural to consider that point as origin. We actually take the proper Lorentz transformations with determinant $+1$ as the local symmetry group.

In the following, we introduce gauge-vector fields of the type considered in Yang-Mills theories in terms of the tetrad fields (Sec.~\ref{secgaugevecs}). Our main objective is to provide a complete set of evolution equations governing both the tetrad and Yang-Mills fields (Sec.~\ref{secevoleqs}). In particular, we propose a new set of coordinate conditions that supplies four missing evolution equations. Based on our complete set of equations, we discuss two applications: planar gravitational waves  (Sec.~\ref{secwaves}) and the exact black-hole solution (Sec.~\ref{secisotropic}). The Hamiltonian formulation of the entire set of evolution equations is an important step for discussing the stability and quantization of the composite gauge theory of gravitation (Sec.~\ref{secHamiltonian}). Finally, we provide a concise summary and discuss the main implications of our results  (Sec.~\ref{secconclusions}). Four appendices cover the structure of the Lorentz group (App.~\ref{AppstructureLG}), the emergence of geometric variables in the Yang-Mills-type theory with Lorentz symmetry group (App.~\ref{appgeovars}), perturbation theory (App.~\ref{secperturb}), and some useful time derivatives (App.~\ref{apptimeder}).

\section{Gauge vector fields}\label{secgaugevecs}
To incorporate the gauge symmetry associated with freely falling local Lorentz frames into a Yang–Mills–type gauge theory, we express the gauge field components $A_{(\kappa\lambda) \rho}$ in terms of the tetrad fields ${b^\kappa}_\mu$ and their spacetime derivatives \cite{hco231},
\begin{eqnarray}
   A_{(\kappa\lambda) \rho} &=&
   \frac{1}{2} \, \mbox{$\bar{b}^\mu$}_{\kappa} \left( \frac{\partial g_{\nu\rho}}{\partial x^\mu}
   - \frac{\partial g_{\mu\rho}}{\partial x^\nu} \right) \mbox{$\bar{b}^\nu$}_{\lambda}
   \nonumber\\
   &+& \frac{1}{2} 
   \left(\frac{\partial b_{\kappa\mu}}{\partial x^\rho} \, \mbox{$\bar{b}^\mu$}_{\lambda}
   - \mbox{$\bar{b}^\mu$}_{\kappa} \, \frac{\partial b_{\lambda\mu}}{\partial x^\rho} 
   \right) , \qquad 
\label{compositionrule}
\end{eqnarray}
where the coefficients $\bar{b}^\mu{}_\kappa$ represent the entries of the inverse of the regular matrix ${b^\kappa}_\mu$. In view of the antisymmetry of this definition in $\kappa$ and $\lambda$, the 16 pairs $(\kappa,\lambda)$ of spacetime indices correspond to the 6 labels for the base vectors of the Lie algebra of the proper Lorentz group (see Table~\ref{tabindexmatch}), so that Eq.~(\ref{compositionrule}) provides the definition of the gauge vector fields $A_{a\rho}$. We refer to Eq.~(\ref{compositionrule}) as the composition rule that expresses the $24$ components of the gauge vector fields in terms of the $16$ tetrad fields, resulting in a severely constrained Yang-Mills-type theory. Note that the gauge vector fields (\ref{compositionrule}) are invariant under a global rescaling of ${b^\kappa}_\mu$.

\begin{table}
\begin{tabular}{c|c c c c c c}
	% \hline
    $a$ \, & \, $1$ & $2$ & $3$ & $4$ & $5$ & $6$ \\
	\hline
	$(\kappa,\lambda)$ \,
    & \, $(0,1)$ & $(0,2)$ & $(0,3)$ & $(2,3)$ & $(3,1)$ & $(1,2)$ \\
	% \hline
\end{tabular}
\caption{Correspondence between the label $a$ for the base vectors of the six-dimensional Lie algebra ${\rm so}(1,3)$ of the proper Lorentz group and ordered pairs $(\kappa,\lambda)$ of spacetime indices.}
\label{tabindexmatch}
\end{table}

From Eqs.~(\ref{gaugetransformb}) and (\ref{Minkowskiinvariance}) we obtain the gauge transformation behavior
\begin{equation}\label{gaugetransformbbar}
   \bar{b}^\mu{}_\kappa \rightarrow {\Lambda_\kappa}^\lambda \; \bar{b}^\mu{}_\lambda \,,
\end{equation}
where indices are lowered or raised by means of the Minkowski metric, and Eqs.~(\ref{gaugetransformb}) and (\ref{gaugetransformbbar}) imply the transformation behavior of gauge-vector fields,
\begin{equation}\label{gaugetransformA}
   A_{(\kappa\lambda) \rho} \rightarrow {\Lambda_\kappa}^{\kappa'} \, {\Lambda_\lambda}^{\lambda'}
   A_{(\kappa'\lambda') \rho} + D_{(\kappa\lambda) \rho} \,,
\end{equation}
with
\begin{equation}\label{derLamdef}
   D_{(\kappa\lambda) \rho} = \eta_{\kappa'\lambda'} \,
   \frac{\partial {\Lambda_\kappa}^{\kappa'}}{\partial x^\rho} \, {\Lambda_\lambda}^{\lambda'} \,.
\end{equation}
The property (\ref{Minkowskiinvariance}) implies that $D_{(\kappa\lambda) \rho}$ is antisymmetric in $\kappa$ and $\lambda$.

For the infinitesimal gauge transformations associated with the Lie algebra ${\rm so}(1,3)$, we write
\begin{equation}\label{infinitesimalGTL}
   {\Lambda_\kappa}^\lambda = {\delta_\kappa}^\lambda + \delta{\Lambda_\kappa}^\lambda .
\end{equation}
From Equation (\ref{Minkowskiinvariance}) it follows that the infinitesimal transformation matrix $\delta\Lambda_{\kappa\lambda}$ is antisymmetric. Using this result, we derive the infinitesimal form of the corresponding finite gauge transformations (\ref{gaugetransformb}), (\ref{gaugetransformbbar}) and (\ref{gaugetransformA}):
\begin{equation}\label{infinitesimalGTb}
   \delta{b^\kappa}_\mu = \delta{\Lambda^\kappa}_\lambda \; {b^\lambda}_\mu \,,
\end{equation}
\begin{equation}\label{infinitesimalGTbbar}
   \delta\bar{b}^\mu{}_\kappa = \delta{\Lambda_\kappa}^\lambda \; \bar{b}^\mu{}_\lambda \,,
\end{equation}
and
\begin{equation}\label{infinitesimalGTa}
   \delta A_{(\kappa\lambda) \rho} = \frac{\partial \delta\Lambda_{\kappa\lambda}}{\partial x^\rho}
   + \eta^{\kappa'\lambda'} \Big[ A_{(\kappa'\lambda) \rho} \, \delta\Lambda_{\kappa\lambda'} 
   - A_{(\kappa\lambda') \rho} \, \delta\Lambda_{\kappa'\lambda} \Big] .
\end{equation}
With the help of Eq.~(\ref{supauxf1}) in App.~\ref{AppstructureLG}, this transformation law for the gauge vector fields can be reformulated in terms of the structure constants of the Lie algebra of the proper Lorentz group.

The expression (\ref{compositionrule}) for the gauge vector fields is severely nonlinear. By transforming from the freely falling frames to the background frame, we obtain a much simpler expression that is linear in $g_{\mu\nu}$ and bilinear in ${b^\kappa}_\mu$,
\begin{eqnarray}\label{Atildef}
   \tilde{A}_{(\mu\nu)\rho} &=& {b^\kappa}_\mu {b^\lambda}_\nu \, A_{(\kappa\lambda) \rho} \\
   &=& \frac{1}{2} \left( \frac{\partial g_{\nu\rho}}{\partial x^\mu}
   - \frac{\partial g_{\mu\rho}}{\partial x^\nu}
   + b_{\kappa\mu} \, \frac{\partial{b^\kappa}_\nu}{\partial x^\rho}
   - b_{\kappa\nu} \, \frac{\partial{b^\kappa}_\mu}{\partial x^\rho} \right) .  \nonumber
\end{eqnarray}
An alternative form of these fields is given by
\begin{equation}\label{Atildefexpl}
   \tilde{A}_{(\mu\nu)\rho} = \frac{1}{2} \left( \frac{\partial g_{\nu\rho}}{\partial x^\mu}
   + \frac{\partial g_{\mu\nu}}{\partial x^\rho}
   - \frac{\partial g_{\mu\rho}}{\partial x^\nu} \right) 
   - {b^\kappa}_\nu \, \frac{\partial b_{\kappa\mu}}{\partial x^\rho} ,  
\end{equation}
where the first term consists of derivatives of the gauge invariant metric. The gauge dependence of $\tilde{A}_{(\mu\nu)\rho}$ results from the last term in Eq.~(\ref{Atildefexpl}). It turns out in the following, that working with various transformed variables is very convenient.

Equation~(\ref{compositionrule}) suggests that the gauge invariant part of the gauge vector fields is closely related to a metric connection. Equation~(\ref{Atildefexpl}) suggests to introduce
\begin{equation}\label{Gammatildef}
   \tilde{\Gamma}_{\rho\mu\nu} \! = \!
   - \tilde{A}_{(\rho\mu)\nu} + {b^\kappa}_\rho \, \frac{\partial b_{\kappa\mu}}{\partial x^\nu} =
   \frac{1}{2} \left( \frac{\partial g_{\rho\nu}}{\partial x^\mu} 
   + \frac{\partial g_{\mu\rho}}{\partial x^\nu}
   - \frac{\partial g_{\nu\mu}}{\partial x^\rho} \right) ,
\end{equation}
which, indeed, is closely related to the Christoffel symbols describing a torsion-free metric connection,
\begin{equation}\label{Gammadef}
   \Gamma^\rho_{\mu\nu} = \bar{g}^{\rho\sigma} \, \tilde{\Gamma}_{\sigma\mu\nu} ,
\end{equation}
where $\bar{g}$ is the inverse of the metric. Note the following useful identities,
\begin{equation}\label{gderivative}
   \frac{\partial g_{\mu\nu}}{\partial x^\rho} =
   \tilde{\Gamma}_{\mu\rho\nu} + \tilde{\Gamma}_{\nu\rho\mu} ,
\end{equation}
and
\begin{equation}\label{bderivative}
   \bar{b}^\rho{}_\kappa \frac{\partial {b^\kappa}_\mu}{\partial x^\nu} =
   \Gamma^\rho_{\mu\nu} + \bar{g}^{\rho\sigma} \tilde{A}_{(\sigma\mu)\nu} .
\end{equation}

From the tetrad variables characterizing freely falling frames, one can naturally construct both a metric and gauge vector fields. The gauge vector fields are in turn closely related to the Christoffel symbols. Yet this is only the tip of the iceberg. As we shall see below, the field tensor of the Yang-Mills theory has the same mathematical structure as the Riemann curvature tensor, and the covariant derivatives of Yang-Mills theory and Riemannian geometry are deeply related (see App.~\ref{appgeovars} for details).

The observation that freely falling frames can lead us both to a Yang-Mills-type theory defined on Minkowski space or to the pseudo-Riemannian geometry of general relativity provides a compelling illustration of Einstein's remark on geometric conventionalism quoted in the beginning of this paper. In the weak-field regime---where Einstein’s celebrated predictions, such as light deflection and gravitational redshift, have been confirmed with high precision---general relativity and the Yang–Mills–type theory are epistemologically equivalent. In contrast, in the strong-field regime---encompassing phenomena such as black holes and the gravitational waves emitted by binary systems of white dwarfs, neutron stars, and black holes---the two theories are found to diverge significantly.

\section{Field equations}\label{secevoleqs}
The evolution equations for the gauge vector fields $A_{(\kappa\lambda) \rho}$ are given by the Yang-Mills-type theory associated with the Lorentz symmetry group. However, the composition rule (\ref{compositionrule}) yields only an incomplete set of evolution equations for the tetrad variables ${b^\kappa}_\mu$. In fact, even the evolution of the gauge-invariant variables $g_{\mu 0}$ remains undetermined by the composition rule. These missing evolution equations must be supplied through so-called coordinate conditions.

Identifying suitable coordinate conditions is the least straightforward step in formulating the composite theory of gravity. Guidance comes from the simplicity of the static isotropic field around a mass point at rest and by aiming at a Hamiltonian structure of the complete set of evolution equations.

\subsection{Yang-Mills equations and conserved currents}\label{subsecJ}
Following standard procedures for Yang-Mills theories (see, e.g., Sect.~15.2 of \cite{PeskinSchroeder}, Chap.~15 of \cite{WeinbergQFT2}, or \cite{hco229}), we can introduce a field tensor $ F_{a \mu\nu}$ in terms of the gauge vector fields,
\begin{equation}\label{Fdefinition}
   F_{a \mu\nu} = \frac{\partial A_{a \nu}}{\partial x^\mu}
   - \frac{\partial A_{a \mu}}{\partial x^\nu}
   + f^{bc}_a A_{b \mu} A_{c \nu} ,
\end{equation}
where $f^{bc}_a$ stands for the structure constants of the Lie algebra, here of ${\rm so}(1,3)$ for the Lorentz group, and the coupling constant is taken as unity. A Lie algebra label, say $a$, can be raised or lowered by raising or lowering the indices of the pairs associated with $a$ (see Table~\ref{tabindexmatch}) by means of the Minkowski metric. The structure constants for ${\rm so}(1,3)$ can then be specified as follows: $f^{abc}$ is $1$ ($-1$) if $(a,b,c)$ is an even (odd) permutation of $(4,5,6)$, $(4,3,2)$, $(5,1,3)$ or $(6,2,1)$, and $0$ otherwise. Note that $F_{a \mu\nu}$ is antisymmetric in $\mu$ and $\nu$.

The field equations of Yang-Mills theories are given by
\begin{equation}\label{YMfieldeqs}
   \frac{\partial F_{a \mu\nu}}{\partial x_\mu} + f_a^{bc} \, A_b^\mu F_{c \mu\nu} = - J_{a \nu} ,
\end{equation}
where the quantities $J_{a \nu}$ characterize the external sources of the field. By acting with the derivative $\partial/\partial x_\nu$ on Eq.~(\ref{YMfieldeqs}) and using the properties of the structure constants, one can derive the conservation laws
\begin{equation}\label{Jaconservation}
   \frac{\partial J_{a \nu}}{\partial x_\nu} + f_a^{bc} \, A_b^\nu J_{c \nu} = 0 .
\end{equation}
We therefore refer to $J_{a \nu}$ as the conserved currents of the Yang-Mills theory.

To express the sources of a gravitational field in terms of the energy-momentum tensor ${T_\mu}^\nu$ we introduce the transformed variables
\begin{equation}\label{Jhatdef}
  \hat{J}^{(\mu\nu)\rho} =
  \mbox{$\bar{b}^\mu$}_{\kappa} \mbox{$\bar{b}^\nu$}_{\lambda} \, J^{(\kappa\lambda)\rho} .
\end{equation}
The following identity for the coupling contribution to the Hamiltonian of Yang-Mills theories suggests that the currents $\hat{J}^{(\mu\nu)\rho}$ are the natural partners of the vector fields $\tilde{A}_{(\mu\nu)\rho}$,
\begin{eqnarray}
   \sum_{a\rho} A_{a\rho} \, J^{a\rho} &=&
   \frac{1}{2} \sum_{\kappa\lambda\rho} A_{(\kappa\lambda)\rho} \, J^{(\kappa\lambda)\rho} =
   \frac{1}{2} \sum_{\mu\nu\rho} \tilde{A}_{(\mu\nu)\rho} \, \hat{J}^{(\mu\nu)\rho} \nonumber \\
   && \hspace{-1.3cm} = \, \sum_{n\rho} \left( \tilde{A}_{(0n)\rho} \, \hat{J}^{(0n)\rho}
   + \tilde{A}_{(\Obar{n}\Ubar{n})\rho} \, \hat{J}^{(\Obar{n}\Ubar{n})\rho} \right) ,
\label{AJ}
\end{eqnarray}
where, for compact notation, the larger and smaller neighbors of a space index $n$ are introduced as $\Obar{n}$ and $\underbar{\it n}$ in Table~\ref{tabneighbors}.

\begin{table}
\setlength{\tabcolsep}{10pt}
\begin{tabular}{ c c c}
    $n$  & $\Obar{n}$ & $\Ubar{n}$ \\
	\hline
	1 & 2 & 3 \\
	2 & 3 & 1 \\
	3 & 1 & 2 \\
\end{tabular}
\caption{Larger and smaller neighbors of a space index $n$ under cyclic boundary conditions.}
\label{tabneighbors}
\setlength{\tabcolsep}{6pt}
\end{table}

The currents $\hat{J}^{(\mu\nu)\rho}$ are gauge invariant, whereas $J^{(\kappa\lambda)\rho}$ must match the gauge transformation behavior of the left-hand side of Eq.~(\ref{YMfieldeqs}) so that the field equations become gauge independent. The currents admit an explicit representation in a transparent form that clearly reveals their underlying structure \cite{hco252},
\begin{equation}\label{JhatfromT}
   \hat{J}^{(\mu\nu)\rho} = \frac{8 \pi G}{c^4} \left( 
   \bar{g}^{\mu\sigma} \frac{{\cal D} \hat{S}^{\nu\rho}}{{\cal D} x^\sigma}
   - \bar{g}^{\nu\sigma} \frac{{\cal D} \hat{S}^{\mu\rho}}{{\cal D} x^\sigma} \right) ,
\end{equation}
where $G$ is Newton's constant, 
% $G=6.67 \, \frac{\rm{m}^3}{\rm{kg}\,\rm{s}^2}$ ,
$c$ is the speed of light, and covariant derivatives of the symmetric tensors
\begin{equation}\label{symTdef}
   \hat{S}^{\mu\nu} = \bar{g}^{\mu\sigma} {T_\sigma}^\nu 
   - \frac{1}{2} \bar{g}^{\mu\nu} \, {T_\sigma}^\sigma ,
\end{equation}
are defined by
\begin{equation}\label{SymTder}
   \frac{{\cal D} \hat{S}^{\mu\nu}}{{\cal D} x^\rho} =
   \frac{\partial \hat{S}^{\mu\nu}}{\partial x^\rho}
   + \Gamma^\mu_{\rho\sigma} \hat{S}^{\sigma\nu}
   + \Gamma^\nu_{\rho\sigma} \hat{S}^{\mu\sigma}
   - \Gamma^\sigma_{\sigma\rho} \, \hat{S}^{\mu\nu} .
\end{equation}

Equation (\ref{JhatfromT}) implies the cyclicity property
\begin{equation}\label{Jhatcyclegen}
   \hat{J}^{(\mu\nu)\rho} + \hat{J}^{(\nu\rho)\mu} + \hat{J}^{(\rho\mu)\nu} = 0 ,
\end{equation}
and, together with energy-momentum conservation, the contraction property
\begin{equation}\label{Jhatgrad}
   \hat{J}^{(\sigma\mu)\rho} g_{\rho\sigma} = \frac{4\pi G}{c^4} \bar{g}^{\mu\nu} \bigg(
   \frac{\partial {T_\rho}^\rho}{\partial x^\nu} - \Gamma^\sigma_{\sigma\nu} {T_\rho}^\rho \bigg) .
\end{equation}
With the help of Eq.~(\ref{supauxf2}), the conservation law (\ref{Jaconservation}) can be rewritten as
\begin{equation}\label{Jhatconservation}
   \frac{\partial \hat{J}^{(\mu\nu)\rho}}{\partial x^\rho}
   + \Gamma^\mu_{\rho\sigma} \hat{J}^{(\sigma\nu)\rho}
   + \Gamma^\nu_{\rho\sigma} \hat{J}^{(\mu\sigma)\rho} = 0 .
\end{equation}
A comparison of the corresponding conservation laws (\ref{Jaconservation}) and (\ref{Jhatconservation}) shows that the covariant derivatives in the gauge theory of gravity and in pseudo-Riemannian geometry are closely related (see App.~\ref{appgeovars} for details).

Evolution equations for the currents $\hat{J}^{(mn)\sigma}$ follow from the structure of the expressions (\ref{JhatfromT}). For simplicity, we only give the linearized equations, which are a special case of another cylicity property,
\begin{equation}\label{Bianchit}
   \frac{\partial \hat{J}^{(mn)\sigma}}{\partial x^0} =
   \frac{\partial \hat{J}^{(0m)\sigma}}{\partial x_n}
   - \frac{\partial \hat{J}^{(0n)\sigma}}{\partial x_m} .
\end{equation}

\subsection{Coordinate conditions}
The problem of missing evolution equations for the metric is familiar from general relativity. It is typically resolved by imposing coordinate conditions on the metric, which can be fulfilled through a nonlinear coordinate transformation. In general relativity, such transformations have no impact on physical predictions (for Einstein's historical struggle with coordinate conditions, see \cite{Giovanelli21}). In contrast, in the composite Yang-Mills theory of gravity, restrictions on the form of the metric have physical significance: they impose genuine constraints on the admissible families of freely falling frames.

A common way of formulating the missing evolution equations in Lorentz covariant form is given by
\begin{equation}\label{coco}
    \frac{\partial g_{\mu\rho}}{\partial x_\rho}
   - K \frac{\partial {g_\rho}^\rho}{\partial x^\mu} = Z \tilde{A}^\rho_{(\rho\mu)} ,
\end{equation}
where $K$ and $Z$ are constants. By linearization of the harmonic coordinate conditions of general relativity \cite{Weinberg},
\begin{equation}\label{cocoharm}
   \bar{g}^{\rho\nu} \left(  \frac{\partial g_{\mu\nu}}{\partial x^\rho}
   - \frac{1}{2} \frac{\partial g_{\rho\nu}}{\partial x^\mu} \right) = 0 ,
\end{equation}
we obtain $K=1/2$, $Z=0$ (used in \cite{hco250}). In \cite{hco252}, the differential operator
$\partial^2/\partial x^\rho \partial x_\rho$
has been applied on the linearized harmonic coordinate conditions to suppress the need for a nonzero right-hand side in Eq.~(\ref{coco}). Less appealing coordinate conditions are implied by the use of Schwarzschild coordinates in \cite{CamenzindCam75,Camenzind77b} or of quasi-Minkowskian coordinates in \cite{hco231}.

In general, the right-hand side of Eq.~(\ref{coco}) is not gauge invariant. According to Eq.~(\ref{Atildefexpl}), gauge invariance is achieved for
\begin{equation}\label{cocof}
    \frac{\partial {b^\kappa}_\mu}{\partial x_\mu} = 0 .
\end{equation}

With the composition rule (\ref{compositionrule}), the coordinate conditions (\ref{coco}) can be rewritten as
\begin{equation}\label{cocos}
    b_{\kappa\mu} \frac{\partial{b^\kappa}_\rho}{\partial x_\rho} +
    \left( 1 - Z \right) b_{\kappa\rho} \frac{\partial {b^\kappa}_\mu}{\partial x_\rho} =
    \left( K - \frac{1}{2} Z \right) \frac{\partial {g_\rho}^\rho}{\partial x^\mu} .
\end{equation}
The Newtonian limit for the gravitational field surrounding a point mass $m$, which is characterized by
\begin{equation}\label{Newtong}
   g_{\mu\nu} = \left( \begin{matrix}
   -1 + \frac{2 G m}{r c^2} & 0 \\
   0 & 1
   \end{matrix} \right) ,
   \quad 
   {b^\kappa}_\mu = \left( \begin{matrix}
   1 - \frac{G m}{r c^2} & 0 \\
   0 & 1
   \end{matrix} \right) ,
\end{equation}
implies
\begin{equation}\label{cocoNewton}
   2 K - Z = 0 .
\end{equation}
The most natural choice is
\begin{equation}\label{natparameters}
    K=1/2, \quad Z=1 .
\end{equation}
In this case, the conditions (\ref{coco}) and (\ref{cocof}) actually turn out to be equivalent. The conditions (\ref{cocof}) provide the missing four time evolution equations for the tetrad variables ${b^\kappa}_0$.

In summary, the fundamental equations of the composite theory of gravity are given by the composition rule (\ref{compositionrule}), the Yang-Mills field equations (\ref{YMfieldeqs}), and the coordinate conditions (\ref{cocof}). The coordinate conditions differ from all previous choices.

\subsection{Gauge fixing conditions}\label{secgaugefix}
In the Yang–Mills theories of electroweak and strong interactions, only half of the degrees of freedom are physical, while the other half is eliminated through appropriate gauge conditions. For instance, for electromagnetic waves propagating in vacuum, the physical degrees of freedom can be represented by two transverse polarization modes, whereas the longitudinal and temporal polarization modes are considered as unphysical. This situation changes in composite Yang–Mills theories, where the gauge vector fields are constructed from a reduced set of more fundamental degrees of freedom from the outset.

The simplicity of the gauge transformation (\ref{gaugetransformb}) for the tetrad variables, compared to the transformation laws (\ref{gaugetransformA}), (\ref{derLamdef}) for the gauge vector fields, actually facilitates the selection of a convenient gauge significantly. A Lorentz transformation can be applied independently at each point in space and time. We assume, however, that these Lorentz transformations vary smoothly in space and time.

A special decomposition (\ref{localinertialdecomp}) of a metric $g_{\mu\nu}$ is obtained by the definition
\begin{equation}\label{squarerootgauge}
   \bar{b}^\mu{}_\kappa = \frac{e^\mu_{(\kappa)}}{\sqrt{|\lambda_{(\kappa)}|}} ,
\end{equation}
where $e_{(\kappa)}$ denotes four orthonormal eigenvectors of the symmetric matrix $g_{\mu\nu}$. As in general relativity, one eigenvalue is negative ($\lambda_{(0)}$) and three are positive ($\lambda_{(1)}$, $\lambda_{(2)}$, $\lambda_{(3)}$). For the $\bar{b}^\mu{}_\kappa$ defined in Eq.~(\ref{squarerootgauge}), one obtains $\bar{b}^\mu{}_\kappa \, g_{\mu\nu} \, \bar{b}^\nu{}_\lambda = \eta_{\kappa\lambda}$, which is equivalent to the decomposition (\ref{localinertialdecomp}). A characteristic feature of this decomposition is that the column vectors of $\bar{b}^\mu{}_\kappa$ are orthogonal, though not normalized. The pairwise orthogonality of four nonzero column vectors implies six independent conditions, corresponding to six gauge constraints. We refer to this particular choice of $\bar{b}^\mu{}_\kappa$ as the `square-root' or `semi-orthogonal' decomposition of the metric $g_{\mu\nu}$.

Since the Lie algebra of the Lorentz group is the six-dimensional space of antisymmetric $4 \times 4$ matrices, the antisymmetric part of ${b^\kappa}_\mu$ can be altered by the gauge transformations (\ref{infinitesimalGTb}). Consequently, imposing that ${b^\kappa}_\mu$ be symmetric provides another natural and convenient gauge. In analogy with Cholesky decompositions, ${b^\kappa}_\mu$ may alternatively be constrained to take an upper or lower triangular form.  More generally, one may distribute six vanishing entries partly above and partly below the diagonal, or impose symmetry conditions on selected components, depending on the symmetries of the problem at hand. In any complete gauge fixing, the number of degrees of freedom in the tetrad variables must be reduced by six.

While a universal gauge is attractive from a conceptual standpoint, specific problems often admit gauge choices that greatly simplify their analysis. In this work, we advocate a gauge‑fixing strategy that combines universal applicability with problem‑dependent practicality. Our aim is to select six gauge conditions for the tetrad variables--chosen from the various options discussed above--in a manner that ensures the covariant Lorenz gauge
\begin{equation}\label{Lorenzgauge}
   \frac{\partial A_{a \mu}}{\partial x_\mu} = 0 .
\end{equation}
As discussed in detail below, for the isotropic static gravitational field surrounding a point mass, a symmetric ${b^\kappa}_\mu$ is consistent with Eq.~(\ref{Lorenzgauge}). In the case of planar gravitational waves propagating in vacuum, the symmetric gauge is inappropriate, but the Lorenz gauge can be achieved by appropriately distributing four zeros above and below the diagonal of  ${b^\kappa}_\mu$ and imposing two symmetry conditions on the antidiagonal. The Lorenz gauge is not an added constraint, but rather the reward for a proper implementation of the gauge on the level of tetrad variables.

The time derivative of the constraints (\ref{Lorenzgauge}) leads to the conditions
\begin{equation}\label{gaugeconstrEa02}
    J^{a0} = \frac{\partial E^a_n}{\partial x_n} + f_c^{ab} A_{b n} E^{c n} ,
\end{equation}
where $E^a_n=F^a_{n0}$ is given by Eq.~(\ref{Fdefinition}). This relation, which constitutes a generalization of Gauss's law for the electric field, follows directly from the field equation (\ref{YMfieldeqs}) for $\nu=0$ without any need to impose the Lorenz gauge. It therefore does not introduce additional constraints.

\subsection{Two potential concerns}
We arrive at a formidable system of evolution equations involving the $16$ tetrad variables ${b^\kappa}_\mu$ and the $24$ components of the gauge vector fields $A^a_\mu$. The tetrad variables satisfy first‑order evolution equations, whereas the gauge vector fields obey the second‑order equations of Yang-Mills theory and are simultaneously constrained by the composition rule. Using the composition rule (\ref{compositionrule}), the field equations (\ref{YMfieldeqs}) can be reformulated as third‑order differential equations for the tetrad variables.

We may worry about the huge number of degrees of freedom and the potential stability issues inherent in higher-derivative theories \cite{Stelle77,Stelle78,Krasnikov87,Beckeretal17,GrosseKnetter94,PaisUhlenbeck50,Ostrogradsky1850,Woodard15,Gitmanetal83,Chenetal13,RaidalVeermae17,hco235,hco237}. These serious concerns are addressed by studying concrete examples in the following two sections and by elaborating the Hamiltonian structure of our system of equations, which is also the key to quantization. In general, a substantial set of constraints eliminates the issues mentioned above.

\section{Planar wave propagation}\label{secwaves}
We now consider planar gravitational waves propagating in vacuum along the $3$-direction. We assume that all variables exhibit a space‑ and time‑dependence of the form $e^{ik (x_3 - c t)}$, where $k>0$ is the wave number. For each variable, a complex prefactor encodes the amplitude and initial phase of the wave.

Adapting the gauge conditions to wave propagation along the $3$-direction,  we set four components of ${b^\kappa}_\mu$ to zero and impose two symmetry conditions along the antidiagonal,
\begin{equation}\label{planarwavesbmax}
   {b^\kappa}_\mu = {\delta^\kappa}_\mu + \left( \begin{array}{cccc}
      \frac{1}{2} h_{00} & 0 & 0 & \frac{1}{2} h_{30} \\
      h_{10} & \frac{1}{2} h_{11} & \frac{1}{2} h_{12} & h_{13} \\
      h_{20} & \frac{1}{2} h_{12} & \frac{1}{2} h_{22} & h_{23} \\
      \frac{1}{2} h_{30} & 0 & 0 & \frac{1}{2} h_{33}
   \end{array} \right) .
\end{equation}
The coordinate conditions (\ref{cocof}) imply
\begin{equation}\label{cocowaves}
   h_{13} = - h_{10} , \quad h_{23} = - h_{20} , \quad h_{33} = h_{00} = - h_{30} ,
\end{equation}
thus further reducing the number of degrees of freedom from ten to six. As in Chapter~10 of \cite{Weinberg} in the context of general relativity, we treat gravitational radiation in the weak-field approximation, that is, to first order in $h_{\mu\nu}$.

The gauge vector fields can be calculated from Eq.~(\ref{planarwavesbmax}) and the composition rule (\ref{compositionrule}),
\begin{equation}\label{Amatrix}
   A^a_\mu = \frac{ik}{2} \left( \begin{array}{cccccc}
      0 & 0 & 0 & 0 & 0 & 0 \\
      h_{11} & h_{12} & 0 & -h_{12} & h_{11} & 0 \\
      h_{12} & h_{22} & 0 & -h_{22} & h_{12} & 0 \\
      0 & 0 & 0 & 0 & 0 & 0
   \end{array} \right)^{\hspace{-.3em} T} ,
\end{equation}
with the properties
\begin{equation}\label{Avectorwaves}
   A^{(01)}_\mu = A^{(31)}_\mu , \quad A^{(02)}_\mu = A^{(32)}_\mu .
\end{equation}
Only two transverse spatial vector fields are required to fully characterize the propagation of planar gravitational waves. The gauge vector fields (\ref{Amatrix}) satisfy the field equations (\ref{YMfieldeqs}).

Temporal and longitudinal modes associated with $h_{l0}$ may arise in the tetrad variables, but they do not affect the gauge vector fields, which remain strictly transverse. General relativity is consistent with the observation that only the central $2 \times 2$ sub-matrix of Eq.~(\ref{planarwavesbmax}) is physically relevant to the gauge vector fields, together with the additional constraint $h_{22}=-h_{11}$ (see p.\,256 of \cite{Weinberg}). This condition implies that the two vectors appearing in Eq.~(\ref{Avectorwaves}) should be orthogonal.  Since the gravitational field couples to matter solely through the conserved currents entering the field equations (\ref{YMfieldeqs}), it is natural to conclude that the three modes associated with $h_{l0}$ in the tetrad variables cannot be excited.

Our gauge choice (\ref{planarwavesbmax}) for the tetrad variables turns out to be successful: it implies the Lorenz gauge (\ref{Lorenzgauge}). From the definition of the field tensor, we further obtain
\begin{equation}\label{Ematrixtri}
   E^a_n = ik A^a_n ,
\end{equation}
which is exactly like the relation between electric field vector and vector potential for electromagnetic waves in the radiation gauge. According to Eq.~(\ref{Avectorwaves}), there are only two independent gauge vector fields. Whereas Weinberg associates gravitational waves with spin $2$ (see p.\,258 of \cite{Weinberg}), the Yang-Mills formulation clearly favors the idea of gravitational waves of spin $1$ associated with the gauge vector fields.

\section{Static isotropic field}\label{secisotropic}
For describing the static isotropic field surrounding a particle resting at the origin we choose the symmetric gauge for the tetrad variable. The isotropy of the field reduces the number of free parameters from ten to three,
\begin{equation}\label{isoxb}
   {b^\kappa}_\mu = \left( \begin{matrix}
   b & 0 \\
   0 & a \left( \delta_{km} - \frac{x_k x_m}{r^2} \right)  + q \frac{x_k x_m}{r^2}
   \end{matrix} \right) ,
\end{equation}
with suitable functions $a$, $b$ and $q$ of the distance $r$ from the origin. The metric (\ref{localinertialdecomp}) is given by
\begin{equation}\label{isoxg}
   g_{\mu\nu} = \left( \begin{matrix}
   -b^2 & 0 \\
   0 & a^2 \left( \delta_{mn} - \frac{x_m x_n}{r^2} \right)  + q^2 \, \frac{x_m x_n}{r^2}
   \end{matrix} \right) .
\end{equation}
At a fixed distance from the origin, freely falling reference frames are connected through orthogonal transformations in space. The function $b(r)$ characterizes the effect of gravitational time dilation. The temporal component ${b^0}_0 = b$ can change sign, whereas $g_{00} = -b^2$ cannot become positive; a value of zero may be reached in strong gravitational fields when $b$ changes sign.

From the coordinate conditions (\ref{cocof}) we obtain
\begin{equation}\label{cocofiso}
   a = q + \frac{1}{2}  r \, q' ,
\end{equation}
where the prime indicates the derivative with respect to $r$. The composition rule (\ref{compositionrule}) leads to
\begin{equation}\label{isoxA}
   A_{(0l)\rho} = \frac{b'}{rq} \, x_l \, \delta_{\rho 0} , \quad 
   A_{(kl)\rho} = \frac{q-a-ra'}{r^2 q} \Big( x_l \delta_{k\rho} - x_k \delta_{l\rho} \Big) .
\end{equation}
These expressions are consistent with the previously given static isotropic solution given in Sec.~V.C of \cite{hco231},
\begin{equation}\label{Aisotropic}
   A_{a\nu} = \frac{r_0}{r^2(r+r_0)} \left(  \begin{matrix}
            x_1 & x_2 & x_3 & 0 & 0 & 0 \\
            0 & 0 & 0 & 0 & -x_3 & x_2 \\
            0 & 0 & 0 & x_3 & 0 & -x_1 \\
            0 & 0 & 0 & -x_2 & x_1 & 0 \\
            \end{matrix} \right)^T ,
\end{equation}
with
\begin{equation}\label{r0def}
   r_0 = \frac{Gm}{c^2} .
\end{equation}
Note that $2 r_0$ is the distance from a large central mass $m$ at the origin at which the Newtonian gravitational potential is strong enough that the classical escape velocity equals the speed of light. This distance is of the same order as the Schwarzschild radius \cite{Weinberg,MisnerThorneWheeler}. Because gravity is fundamentally geometric, the solution is fully characterized by the single length scale $r_0$.

By equating the coefficients in Eqs.~(\ref{isoxA}) and (\ref{Aisotropic}), we find first-order differential equations for $a$ and $b$, which complement the differential equation (\ref{cocofiso}) for $q$ arising from the coordinate conditions. In contrast to the solutions for the coordinate conditions used in previous publications, we here obtain the simple analytical solutions
\begin{equation}\label{abqNEW}
   a = 1 + \frac{1}{2}\frac{r_0}{r} , \quad b = 1 - \frac{r_0}{r} , \quad q = 1 + \frac{r_0}{r} ,
\end{equation}
which contain no higher‑order terms in $1/r$. Given the simplicity of the gauge vector fields in Eq.~(\ref{Aisotropic}), which arise from the tetrad variables, it is reasonable to expect that an appropriate choice of coordinate conditions will lead to an equally simple functional form for the tetrad variables themselves. Note also that, with our adoption of the symmetric gauge, we again succeed in securing the Lorenz gauge (\ref{Lorenzgauge}). The absence of $1/r^2$ corrections to $b$ is essential for reproducing the high-precision predictions of general relativity.

\begin{figure}
\centerline{\includegraphics[width=8 cm]{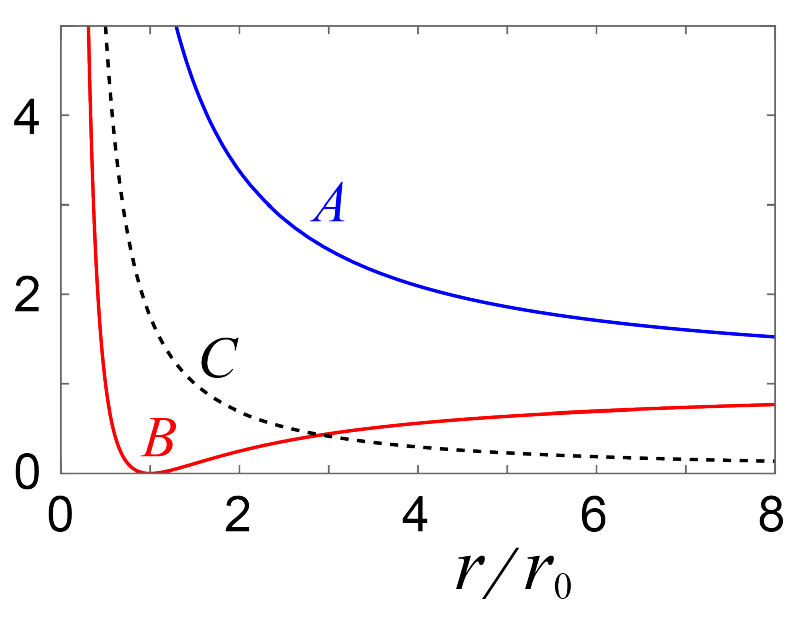}}
\caption[ ]{Exact solutions (\ref{ABCNEW}) for the functions $A$, $B$ and $C$ characterizing the static isotopic metric according to the definitions (\ref{ABCdef}).} \label{plot_analytical_ABC}
\end{figure}

For ease of comparison with previous work, we parameterize the metric in terms of the following three functions of the radial distance $r$,
\begin{eqnarray}
   A &=& g_{nn} - 2 , \nonumber\\
   B &=& - g_{00} , \nonumber\\
   C &=& \frac{r^2}{x_m x_n} \, g_{mn} \quad  \mbox{for } m \neq n .
\label{ABCdef}
\end{eqnarray}
The definitions of the functions $A$ and $B$ for the diagonal components of the metric coincide with the ones used in \cite{hco231}. The function $C$ characterizes the off-diagonal components of the isotropic metric. The quasi-Minkowskian coordinates of \cite{hco231} imply $C=A-1$. For the metric (\ref{isoxg}), the functions defined in (\ref{ABCdef}) are given by
\begin{equation}\label{abq2ABC}
   A = 2(a^2-1) + q^2 , \quad B = b^2 , \quad  C = q^2 - a^2 ,
\end{equation}
or, more explicitly, by
\begin{equation}\label{ABCNEW}
   A = 1 + 4 \frac{r_0}{r} + \frac{3}{2} \frac{r_0^2}{r^2}, \quad
   B = \left( 1 - \frac{r_0}{r} \right)^2 \!\!\! , \quad
   C = \frac{r_0}{r} + \frac{3}{4} \frac{r_0^2}{r^2} .
\end{equation}
These functions are shown in Fig.~\ref{plot_analytical_ABC}. We obtain a simple closed-form black-hole solution that is free of singularities for $r>0$. This observation is essential because, unlike in general relativity, singularities in composite gravity cannot be eliminated through general coordinate transformations. The only peculiarity is that time comes to a standstill at the Schwarzschild radius $r_0$. One might interpret $b<0$ for $r<r_0$ as time running backward inside a black hole. Note that $B = - g_{00}$ is insensitive to the sign of $b$, so a reversal of the time direction cannot be detected from the metric.

\section{Hamiltonian formulation}\label{secHamiltonian}
In this section, we develop composite gravity from the viewpoint of the canonical Hamiltonian approach. This framework offers three major advantages: (i) it clarifies the structure of the constraints and enables a precise count of the theory’s degrees of freedom \cite{hco240}; (ii) it facilitates the analysis and control of Ostrogradsky instabilities \cite{Ostrogradsky1850,Woodard15} in higher derivative theories; and (iii) it provides the foundation for quantizing the theory. Moreover, the Hamiltonian approach guarantees energy conservation and serves as the natural starting point for generalizations to dissipative systems. In particular, this approach allows the formulation of quantum master equations \cite{BreuerPetru,Weiss,hco199,hco221} and renders gravity compatible with the robust framework of dissipative quantum field theory \cite{hcoqft,hco243}.

A successful Hamiltonian formulation of a theory depends critically on the appropriate choice of canonical variables. Factorizing the metric has long been a powerful strategy in this context \cite{Ashtekar86,Ashtekar87}. Whereas one typically limits the number of degrees of freedom by selecting a small symmetry group associated with the decomposition of the metric, composite gravity is built on the six‑parameter Lorentz group supplemented by an additional selection principle that identifies the physically relevant solutions. This ensures that the number of physical degrees of freedom remains appropriately small for a viable theory of gravity.

\subsection{Canonical variables and their evolution}
Following the general ideas for the natural Hamiltonian formulation of composite theories \cite{hco237}, the configurational variables are given by the gauge vector fields $A_{a \mu}$ of the Yang-Mills theory based on the Lorentz group and the tetrad variables ${b^\kappa}_\mu$ used in the composition rule (\ref{compositionrule}) for constructing the gauge vector fields. The respective conjugate momenta are denoted by $E^{a \mu}$ and ${p_\kappa}^\mu$ where, for dimensional consistency, it is convenient to express conjugate momenta in units of the reduced Planck constant $\hbar$. In the generalization (\ref{gaugeconstrEa02}) of Gauss's law, the conjugate momentum associated with the gauge vector fields $A_{am}$ have already been introduced as suitable components of the field tensor,
\begin{equation}\label{Eamdef}
   E^a_n = F^a_{n0} .
\end{equation}
We further introduce the conjugate momentum associated with $A_{a0}$ as
\begin{equation}\label{Ea0def}
   E^a_0 = \frac{\partial A^a_\mu}{\partial x_\mu} .
\end{equation}
For $E^a_0=0$, we recover the Lorenz gauge (\ref{Lorenzgauge}). However, we rather consider $E^a_0$ as a dynamic variable. In contrast to the gauge vector fields $A_{a \mu}$, their conjugate momenta $E^{a \mu}$ are not four-vector fields. By definition, the variable $E^{a 0}$ is a Lorentz scalar, whereas the variables $E^{a m}$ are components of the field tensor.

The physical interpretation of the conjugate momenta ${p_\kappa}^\mu$ in terms of conserved currents follows from the evolution equations for the tetrad variables implied by the composition rule and the coordinate conditions. To clarify the details, we need to determine the Hamiltonian that governs the evolution of the tetrad variables.

The canonical Hamiltonian evolution equations for the tetrad variables and their conjugate momenta are given by
\begin{equation}\label{Hamtetradeqs}
   \frac{\partial {b^\kappa}_\mu}{\partial t} =
   \frac{1}{\hbar} \frac{\delta H}{\delta {p_\kappa}^\mu} ,
   \qquad
   \frac{\partial {p_\kappa}^\mu}{\partial t} =
   - \frac{1}{\hbar} \frac{\delta H}{\delta {b^\kappa}_\mu} ,
\end{equation}
where $H$ is the total Hamiltonian. The evolution equations for the Yang-Mills fields are obtained from the following equations (note the unusual sign conventions),
\begin{equation}\label{HamYMeqs}
   \frac{\partial A_{a \mu}}{\partial t} = - \frac{1}{\hbar} \frac{\delta H}{\delta E^{a \mu}} ,
   \qquad
   \frac{\partial E^{a \mu}}{\partial t} = \frac{1}{\hbar} \frac{\delta H}{\delta A_{a \mu}} .
\end{equation}

As noted earlier, Planck's constant is introduced for dimensional reasons. The necessity of this step follows from Einstein's equivalence principle, which motivates the geometric formulation of gravity, expressed either through the quasi-Riemannian geometry of general relativity or through the differential‑geometric framework underlying Yang–Mills theories. An analogous situation appears in classical equilibrium thermodynamics, where the logarithmic form of the entropy of an ideal gas requires the introduction of an action constant on purely dimensional grounds (see, for example, Sec.~7.3 of \cite{Reif} or Sec.~9.D.3 of \cite{Reichl}). Quantum statistical mechanics reveals the deeper reason for the occurrence of Planck's constant. Likewise, the Hamiltonian formulation of gravity suggests that a complete understanding of gravitational phenomena must ultimately rest on quantum field theory. Newton's constant $G$ appears most naturally in the definition of geometry-related fluxes in terms of the energy-momentum tensor in Eq.~(\ref{JhatfromT}).

\subsection{Exploitation of composition rule}\label{subsecexploicomp}
\subsubsection*{Constraints}
By inserting the gauge vector fields defined in the composition rule (\ref{Atildefexpl}), one can verify the identities
\begin{equation}\label{primconstra}
   \tilde{A}_{(mn)j} - {b^\kappa}_m \, \frac{\partial b_{\kappa n}}{\partial x^j} =
   - \frac{1}{2} \left( \frac{\partial g_{mj}}{\partial x^n} +
   \frac{\partial g_{mn}}{\partial x^j} - \frac{\partial g_{nj}}{\partial x^m} \right) ,
\end{equation}
and
\begin{equation}\label{primconstrb}
   \tilde{A}_{(0m)n} - {b^\kappa}_0 \, \frac{\partial b_{\kappa m}}{\partial x^n} =
   \tilde{A}_{(0n)m} - {b^\kappa}_0 \, \frac{\partial b_{\kappa n}}{\partial x^m} .
\end{equation}
Note that Eqs.~(\ref{primconstra}) and (\ref{primconstrb}) involve only spatial derivatives and no time derivatives. Therefore, they represent constraints on the fields resulting from the composition rule. We refer to them as primary constraints of the composite theory. Eq.~(\ref{primconstra}) represents nine constraints, whereas Eq.~(\ref{primconstrb}) provides three constraints.

\subsubsection*{Evolution equations for tetrad fields}
To characterize the evolution equations for the tetrad variables, we introduce the quantities
\begin{equation}\label{Xmunu}
   X_{\mu\nu} = b_{\kappa\mu} \frac{\partial {b^\kappa}_\nu}{\partial x^0} ,
\end{equation}
which satisfy the convenient identity
\begin{equation}\label{gevoleq}
   \frac{\partial g_{\mu\nu}}{\partial x^0} = X_{\mu\nu} + X_{\nu\mu} .
\end{equation}
Only the antisymmetric part of $X_{\mu\nu}$ depends on the specific choice of the decomposition (\ref{localinertialdecomp}) of the metric, that is, on the gauge. Equation (\ref{Gammatildef}) implies the alternative representation
\begin{equation}\label{XAGammaID}
   X_{\mu\nu} =  \tilde{A}_{(\mu\nu)0} + \tilde{\Gamma}_{\mu\nu 0} .
\end{equation}

By evaluating $\tilde{A}_{(mn)0} + \tilde{A}_{(0n)m}$ and $\tilde{A}_{(0n)0} + \tilde{A}_{(mn)m}/2$ with the composition rule, we find the explicit expressions
\begin{equation}\label{Xeqsmn}
   X_{mn} = \tilde{A}_{(mn)0} + \tilde{A}_{(0n)m} + \frac{\partial g_{0m}}{\partial x^n}
   - b_{\kappa 0} \frac{\partial {b^\kappa}_n}{\partial x^m} , 
\end{equation} 
\begin{eqnarray}\label{Xeqs0nH}
   X_{0n} &=& \tilde{A}_{(0n)0} + \frac{1}{2} \tilde{A}_{(mn)m} \\
   &+& \frac{1}{2} \bigg( \frac{\partial g_{00}}{\partial x^n}
   + \frac{1}{2} \frac{\partial g_{mm}}{\partial x^n}
   - b_{\kappa m} \frac{\partial {b^\kappa}_n}{\partial x_m} \bigg) . \nonumber
\end{eqnarray}
These twelve evolution equations follow directly from the composition rule (\ref{Atildefexpl}). They are not unique, since the constraints (\ref{primconstra}), (\ref{primconstrb}) allow modifications to their right-hand sides. The particular choice of these right-hand sides determines how the conjugate momenta of the tetrad variables contribute to the various currents, as discussed below.

The coordinate conditions (\ref{cocof}) lead to the simple representation $X_{\mu 0} = b_{\kappa\mu} \, \partial {b^\kappa}_n / \partial x_n$. In an attempt to relate all conjugate momenta to conserved currents, we propose the following modification of $X_{m0}$, making use of the constraints (\ref{primconstra}),
\begin{eqnarray}
   X_{m0} &=& - \frac{1}{2} \left( \tilde{A}_{(m\Obar{m})\Obar{m}}
   + \tilde{A}_{(\Ubar{m}m)\Ubar{m}} \right)
   + \frac{1}{4} \frac{\partial (g_{\dObar{m}}-g_{\dUbar{m}})}{\partial x^m} \nonumber \\
   &+& b_{\kappa m} \frac{\partial {b^\kappa}_n}{\partial x_n}
   - \frac{1}{2} b_{\kappa\Obar{m}} \frac{\partial {b^\kappa}_m}{\partial x^{\Obar{m}}}
   + \frac{1}{2} b_{\kappa\Ubar{m}} \frac{\partial {b^\kappa}_m}{\partial x^{\Ubar{m}}} ,  
\label{Xeqsm0H}
\end{eqnarray}
but we keep
\begin{equation}\label{Xeqs00H}
   X_{00} = b_{\kappa 0} \frac{\partial {b^\kappa}_n}{\partial x_n}  .
\end{equation}

\subsection{Hamiltonian}\label{subsecHam}
We express the total Hamiltonian of the composite theory of gravity as the sum of three contributions,
\begin{equation}\label{Ham}
   H = H_{\rm YM} + H_{\rm tetrad} + H_{\rm source} ,
\end{equation}
which will be defined and discussed in the following. For the Yang–Mills contribution $H_{\rm YM}$, we employ the standard Hamiltonian, supplemented by a gauge‑fixing term corresponding to the Feynman gauge (see, e.g., Sec.~2.2 of \cite{BassettoNardelliSoldati}, Chap.~15 of \cite{WeinbergQFT2}, or \cite{hco229}),
\begin{eqnarray}
   H_{\rm YM} &=& \hbar c \int \bigg[ \frac{1}{2} E^{a \mu} E_{a \mu}
   + \frac{1}{4} F_{a mn} F^{a mn} - E^{a 0} \frac{\partial A_{a n}}{\partial x_n} \nonumber \\
   &-& E^{a n} \left( \frac{\partial A_{a 0}}{\partial x^n}
   + f_a^{bc} A_{b n} A_{c 0} \right) \bigg] \, d^3 x . \qquad 
\label{Hfield}
\end{eqnarray}

The contribution associated with the tetrad variables is chosen as
\begin{equation}\label{Hamtetrad}
   H_{\rm tetrad} = \hbar \int \frac{\partial {b^\kappa}_\mu}{\partial t} \, {p_\kappa}^\mu \, d^3 x
   = \hbar c \int X_{\mu\nu} \, \hat{p}^{\mu\nu} \, d^3 x ,
\end{equation}
so that Eq.~(\ref{Hamtetradeqs}) reproduces the evolution equations for the tetrad variables. The reformulation in terms of the variables $X_{\mu\nu}$ and the modified conjugate momenta
\begin{equation}\label{phatdef}
   \hat{p}^{\mu\nu} = \bar{b}^{\mu\kappa} {p_\kappa}^\nu ,
\end{equation}
is more convenient for most practical calculations. From Eqs.~(\ref{HamYMeqs}) and (\ref{Xmunu}) we obtain
\begin{equation}\label{bevolX}
   X_{\mu\nu} = \frac{1}{\hbar c}  b_{\kappa\mu} \frac{\delta H}{\delta {p_\kappa}^\nu} =
   \frac{1}{\hbar c} \frac{\delta H}{\delta \hat{p}^{\mu\nu}} .
\end{equation}

We decompose the Hamiltonian (\ref{Hamtetrad}) into parts that depend on the gauge-vector fields and on the tetrad fields, 
\begin{eqnarray}
   H_{\rm tetrad}^A &=& - \hbar c \int \bigg\{ - \tilde{A}_{(0n)\mu} \, \hat{p}^{\mu n}
   + \tilde{A}_{(\Obar{n}\Ubar{n})0} \, ( \hat{p}^{\Ubar{n}\Obar{n}} - \hat{p}^{\Obar{n}\Ubar{n}} )
   \quad \nonumber \\ && \hspace{-3em}
   + \; \tilde{A}_{(\Obar{n}\Ubar{n})\Ubar{n}} \, \frac{\hat{p}^{\Obar{n}0}+\hat{p}^{0\Obar{n}}}{2}
   + \tilde{A}_{(\Obar{n}\Ubar{n})\Obar{n}} \, \frac{\hat{p}^{\Ubar{n}0}-\hat{p}^{0\Ubar{n}}}{2}
   \bigg\} \, d^3 x , 
\label{HamtetradAnl}
\end{eqnarray}
and $H_{\rm tetrad}^B = H_{\rm tetrad} - H_{\rm tetrad}^A$. An analogous decomposition is introduced for the quantities $X_{\mu\nu}$, which we split into $X_{\mu\nu}^A$ and $X_{\mu\nu}^B$.

For the contribution $H_{\rm source}$ to the full Hamiltonian (\ref{Ham}), we assume a functional of the general form 
\begin{equation}\label{Hamsourceform}
   H_{\rm source} = H_{\rm source}[{b^\kappa}_\mu, A_{a \mu}, {T_\mu}^\nu ] .
\end{equation}
This contribution must not depend on the conjugate momenta ${p_\kappa}^\mu$ or $E^{a\mu}$  because the carefully selected evolution equations for the tetrad fields and the evolution equations for the gauge vector fields derived from the composition rule should not be altered. It may, however, depend on the externally prescribed energy-momentum tensor ${T_\mu}^\nu$ as the source of gravity. Note that $H_{\rm source}$ affects the evolution of the variables ${p_\kappa}^\mu$ and the contributions to the conserved currents.

By combining Eqs.~(\ref{Hamtetradeqs}), (\ref{Xmunu}) and (\ref{phatdef}) we obtain the evolution equations
\begin{eqnarray}\label{phatevol2}
   \frac{\partial \hat{p}^{\mu\nu}}{\partial x^0} &=&
   \bar{g}^{\mu\alpha} \Big[ X_{\alpha\beta} \big( \hat{p}^{\nu\beta} - \hat{p}^{\beta\nu} \big)
   + \tilde{A}_{(\alpha\beta)\rho} \, \hat{\jmath}^{(\nu\beta)\rho} \Big] \\
   &-& \bar{b}^{\mu\kappa} \frac{\delta}{\delta {b^\kappa}_\nu}
   \left( \frac{H_{\rm source}}{\hbar c} + \left. \int X_{\alpha\beta}^B \, \hat{p}^{\alpha\beta}
   \, d^3 x \right|_{\hat{p}^{\alpha\beta}} \right) , \qquad \nonumber
\end{eqnarray}
with the definitions
\begin{equation}\label{j0nmuexpl}
  \hat{\jmath}^{(0n)\mu} = - \hat{p}^{\mu n} ,
\end{equation}
\begin{equation}\label{jmn0expl}
   \hat{\jmath}^{(\Obar{n}\Ubar{n})0} = \hat{p}^{\Ubar{n}\Obar{n}} - \hat{p}^{\Obar{n}\Ubar{n}}  ,
\end{equation}
\begin{equation}\label{jmnnexpl}
   \hat{\jmath}^{(\Obar{n}\Ubar{n})\Ubar{n}} = \frac{\hat{p}^{\Obar{n}0}+\hat{p}^{0\Obar{n}}}{2} ,
\end{equation}
\begin{equation}\label{jmnmexpl}
   \hat{\jmath}^{(\Obar{n}\Ubar{n})\Obar{n}} = \frac{\hat{p}^{\Ubar{n}0}-\hat{p}^{0\Ubar{n}}}{2} ,
\end{equation}
and
\begin{equation}\label{jmnlexpl}
   \hat{\jmath}^{(\Obar{n}\Ubar{n})n} = 0 .
\end{equation}
These definitions imply the cyclicity property
\begin{equation}\label{jhatcyclic0}
   \hat{\jmath}^{(\mu\nu)\rho} + \hat{\jmath}^{(\nu\rho)\mu} + \hat{\jmath}^{(\rho\mu)\nu} = 0 ,
\end{equation}
and the contraction property
\begin{equation}\label{jhatsum}
   \hat{j}^{(\mu n)}_\mu = 0 .
\end{equation}

Three steps are essential for establishing the Hamiltonian structure of composite gravity: (i) selecting appropriate coordinate conditions; (ii) using the constraints to find suitable evolution equations for the tetrad variables; and (iii) specifying the functional form of the Hamiltonian $H_{\rm source}$. At this point, our remaining task is to choose $H_{\rm source}$ such that the properties of the conjugate momenta $\hat{p}^{\mu\nu}$ are consistent with the properties (\ref{Jhatcyclegen})--(\ref{Bianchit}) of the conserved currents.

\subsection{Evolution equations for Yang-Mills fields}
The Hamiltonian evolution equations (\ref{HamYMeqs}) yield
\begin{equation}\label{Aevola0}
   \frac{\partial A_{a 0}}{\partial x^0} = - E_{a 0} + \frac{\partial A_{a n}}{\partial x_n} ,
\end{equation}
\begin{equation}\label{Aevolaj}
   \frac{\partial A_{a n}}{\partial x^0} = - E_{a n} + \frac{\partial A_{a 0}}{\partial x^n}
   + f_a^{bc} A_{b n} A_{c 0} ,
\end{equation}
\begin{equation}\label{Eevola0}
   \frac{\partial E^a_0}{\partial x^0} = - \frac{\partial E^a_n}{\partial x_n}
   - f_c^{ab} A_{b n} E^{c n} - j^a_0
   - \frac{1}{\hbar c} \frac{\delta H_{\rm source}}{\delta A_a^0} ,
\end{equation}
\begin{eqnarray}
   \frac{\partial E^a_n}{\partial x^0} &=& - f_c^{ab} A_{b 0} E^c_n 
   -\frac{\partial F^a_{mn}}{\partial x_m} - f_c^{ab} \, A_b^m F^c_{mn} \nonumber\\
   &-& \frac{\partial E^a_0}{\partial x^n} - j^a_n
   - \frac{1}{\hbar c} \frac{\delta H_{\rm source}}{\delta A_a^n} ,
\label{Eevolaj}
\end{eqnarray}
where the components $F_{amn}$ of the field tensor (\ref{Fdefinition}) involve only spatial derivatives and the contributions $j^a_\rho$ to the conserved currents are given by
\begin{equation}\label{jfromjhat}
   j^{(\kappa\lambda)}_\rho = {b^\kappa}_\mu {b^\lambda}_\nu \, \hat{\jmath}^{(\mu\nu)}_\rho .
\end{equation}

Equations (\ref{Aevola0}) and (\ref{Aevolaj}) confirm the definitions of the conjugate momenta $E_{a\mu}$ anticipated in Eqs.~(\ref{Eamdef}) and (\ref{Ea0def}). The further evolution equations (\ref{Eevola0}) and (\ref{Eevolaj}) reproduce the field equations (\ref{YMfieldeqs}) with
\begin{equation}\label{Jarhoid}
   J^{a\rho} = j^{a\rho} - \frac{1}{\hbar c} \frac{\delta H_{\rm source}}{\delta A_{a\rho}}
   + \frac{\partial E^a_0}{\partial x_\rho} .
\end{equation}
This equation links the conserved currents $J^{a\rho}$ to the quantities $j^{a\rho}$, which are in turn defined in terms of the conjugate momenta $\hat{p}^{\mu\nu}$, and the fields $E^a_0$. In what follows, we impose the Lorenz gauge, so that the last term in Eq.~(\ref{Jarhoid}) vanishes (in the spirit of the discussion in Sec.~\ref{secgaugefix}). The conjugate momenta enter the evolution equations in the role of currents, but their relation to the external sources has not yet been established. To complete the construction, we must determine $H_{\rm source}$ so that the conserved currents (\ref{Jarhoid}) for $E^a_0=0$ exhibit all the properties summarized in Sec.~\ref{subsecJ}.

\subsection{Conjugate momenta $\hat{p}^{\mu\nu}$ and conserved currents}
According to Eq.~(\ref{j0nmuexpl}), the twelve currents $\hat{\jmath}^{(0n)\mu}$ are directly determined by the twelve conjugate momenta $\hat{p}^{\mu n}$.  The three currents $\hat{\jmath}^{(\Obar{n}\Ubar{n})0}$ appearing in Eq.~(\ref{jmn0expl}) do not involve any extra degrees of freedom, since they are fixed by the cylicity property (\ref{jhatcyclic0}), which is consistent with the corresponding cyclicity property (\ref{Jhatcyclegen}) of the sources.

In Eqs.~(\ref{jmnnexpl}) and (\ref{jmnmexpl}), the six additional currents $\hat{\jmath}^{(\Obar{n}\Ubar{n})\Ubar{n}}$ and $\hat{\jmath}^{(\Obar{n}\Ubar{n})\Obar{n}}$ are associated with three further conjugate momenta, $\hat{p}^{n0}$. The remaining three degrees of freedom of these currents are removed by the contraction property (\ref{jhatsum}); however, this condition is not fully consistent with the corresponding property (\ref{Jhatgrad}) of the sources. This mismatch must therefore be compensated by an appropriate contribution to the Hamiltonian $H_{\rm source}$. For simplicity and clarity, we illustrate the procedure only for the linearized theory, for which Eq.~(\ref{Jhatgrad}) reduces to
\begin{equation}\label{Jhatgradlin}
   \hat{J}^{(\mu\nu)}_\mu = \frac{4\pi G}{c^4} \frac{\partial {T_\rho}^\rho}{\partial x_\nu}
   = \hat{q}^\nu ,
\end{equation}
where ${T_\rho}^\rho/c^2$ is the rest‑mass density, serving as a scalar source of the gravitational field. The nonzero right‑hand side arises from the Hamiltonian
\begin{equation}\label{Hsource1}
   H^{(1)}_{\rm source} = - \frac{1}{2} \hbar c \int  \, \tilde{A}_{(mn)m} \, \hat{q}^n \, d^3 x .
\end{equation}
Effectively, this contribution to the Hamiltonian replaces $\hat{p}^{0n}$ in Eqs.~(\ref{jmnnexpl}) and (\ref{jmnmexpl}) by $\hat{p}^{0n}-\hat{q}^n$ to obtain the full currents satisfying Eq.~(\ref{Jhatgradlin}). Equation (\ref{Jarhoid}) for the conserved currents can now be rewritten as
\begin{equation}\label{jJ1}
  \hat{J}^{(0n)\mu} = \hat{\jmath}^{(0n)\mu} , \quad 
  \hat{J}^{(\Obar{n}\Ubar{n})0} = \hat{\jmath}^{(\Obar{n}\Ubar{n})0}  ,
\end{equation}
\begin{equation}\label{jJ2}
   \hat{J}^{(\Obar{n}\Ubar{n})\Ubar{n}} = \hat{\jmath}^{(\Obar{n}\Ubar{n})\Ubar{n}}
   - \frac{1}{2} \hat{q}^{\Obar{n}} , \quad 
   \hat{J}^{(\Obar{n}\Ubar{n})\Obar{n}} = \hat{\jmath}^{(\Obar{n}\Ubar{n})\Obar{n}}
   + \frac{1}{2} \hat{q}^{\Ubar{n}} .
\end{equation}
With the definitions
\begin{equation}\label{Phatdef}
   \hat{P}^{\mu n} = - \hat{J}^{(0n)\mu} , \qquad 
   \hat{P}^{n0} = \hat{J}^{(\Ubar{n}n)\Ubar{n}} - \hat{J}^{(\Obar{n}n)\Obar{n}} ,
\end{equation}
where the conserved currents $\hat{J}^{(\mu\nu)\rho}$ in terms of the energy-momentum tensor and the metric are defined in Eq.~(\ref{JhatfromT}), we find the even simpler relations between source terms and conjugate momentum variables,
\begin{equation}\label{pPid}
   \hat{P}^{\mu n} = \hat{p}^{\mu n} , \qquad \hat{P}^{n0} = \hat{p}^{n0} .
\end{equation}

Finally, the three source terms $\hat{J}^{(23)1}$,  $\hat{J}^{(31)2}$ and $\hat{J}^{(12)3}$, that is $\hat{J}^{(\Obar{n}\Ubar{n})n}$, must be generated entirely by $H_{\rm source}$. We therefore introduce the additional contribution
\begin{equation}\label{Hsource2}
   H^{(2)}_{\rm source} =
   - \hbar c \int \tilde{A}_{(\Obar{n}\Ubar{n})n} \, \hat{J}^{(\Obar{n}\Ubar{n})n} \, d^3 x .
\end{equation}
In general, the currents (\ref{JhatfromT}) depend on the energy-momentum tensor ${T_\mu}^\nu$ and, through the metric, also on the tetrad variables. In the linearized theory, however, the dependence of the currents $\hat{J}^{(\mu\nu)\rho}$ on the tetrad variables can be neglected.

After establishing the relation between conserved currents and conjugate momenta, we now list the evolution equations for $\hat{P}^{\mu\nu}$ that must be reproduced for $\hat{p}^{\mu\nu}$ by choosing the final contribution $H^{(3)}_{\rm source}$ appropriately. The conservation laws (\ref{Jhatconservation}) of the linearized theory can be re-written as
\begin{equation}\label{P0nevol}
   \frac{\partial \hat{P}^{0n}}{\partial x^0} = - \frac{\partial \hat{P}^{mn}}{\partial x^m} ,
\end{equation}
and
\begin{eqnarray}
   \frac{\partial (\hat{P}^{\Obar{n}\Ubar{n}}-\hat{P}^{\Ubar{n}\Obar{n}})}{\partial x^0} &=&
   \frac{\partial \hat{J}^{(\Obar{n}\Ubar{n})n}}{\partial x^n}
   + \frac{1}{2} \frac{\partial (\hat{P}^{\Ubar{n}0}-\hat{P}^{0\Ubar{n}})}{\partial x^{\Obar{n}}} 
   \nonumber \\
   &+& \frac{1}{2} \frac{\partial (\hat{P}^{\Obar{n}0}+\hat{P}^{0\Obar{n}})}{\partial x^{\Ubar{n}}} .
\label{Pmnasymevol}
\end{eqnarray}
From the additional evolution equations (\ref{Bianchit}), we obtain
\begin{equation}\label{Pn0evol}
   \frac{\partial \hat{P}^{n0}}{\partial x^0} =
   \frac{\partial (\hat{P}^{\dObar{n}}-\hat{P}^{\dUbar{n}})}{\partial x^n}
   + \frac{\partial \hat{P}^{\Ubar{n}n}}{\partial x^{\Ubar{n}}}
   - \frac{\partial \hat{P}^{\Obar{n}n}}{\partial x^{\Obar{n}}} .
\end{equation}

While the sources $\hat{J}^{(\mu\nu)\rho}$ are third-order tensors in the background Minkowski space, the conjugate momenta $\hat{p}^{\mu\nu}$ are not tensor components. The simplicity of the evolution equations (\ref{P0nevol})--(\ref{Pn0evol}) arises from the fact that they express conservation laws and cylicity properties of the currents. In contrast, the missing evolution equations for the symmetric part $\hat{P}^{\Obar{n}\Ubar{n}}+\hat{P}^{\Ubar{n}\Obar{n}}$ cannot be derived from covariant equations for the currents. Consequently, both these evolution equations and the corresponding initial conditions depend strongly on the choice of the background frame. For example, in Sec.~\ref{secisotropic} we determined the static isotropic field surrounding a particle at rest. In a moving background frame, one would have to solve a fully dynamical problem involving all currents and all conjugate momenta of the tetrad variables.

For a Hamiltonian formulation we must assume that evolution equations for the symmetric part $\hat{p}^{\Obar{n}\Ubar{n}}+\hat{p}^{\Ubar{n}\Obar{n}}$ can be formulated that are independent of the initial conditions. In view of the covariant field equations presented in Sec.~\ref{secevoleqs}, this is a reasonable assumption. The construction of our contribution $H^{(3)}_{\rm source}$ to the Hamiltonian can only rely on reproducing Eqs.~(\ref{P0nevol})--(\ref{Pn0evol}). Ensuring that the evolution of $\hat{p}^{\Obar{n}\Ubar{n}}+\hat{p}^{\Ubar{n}\Obar{n}}$ coincides with that of $\hat{P}^{\Obar{n}\Ubar{n}}+\hat{P}^{\Ubar{n}\Obar{n}}$ requires six additional constraints.

We finally propose a contribution to $H_{\rm source}$ that depends only on the tetra variables and the external sources,
\begin{eqnarray}
   H^{(3)}_{\rm source} &=& \hbar c \int \bigg\{
   b_{\kappa n} \frac{\partial {b^\kappa}_0}{\partial x_n}
   (\hat{P}^{\dObar{n}} - \hat{P}^{\dUbar{n}})
   \nonumber \\
   &+& b_{\kappa n} \frac{\partial {b^\kappa}_0}{\partial x^{\Ubar{n}}} \hat{P}^{\Ubar{n}n}
   - b_{\kappa n} \frac{\partial {b^\kappa}_0}{\partial x^{\Obar{n}}} \hat{P}^{\Obar{n}n}
   - \underline{\frac{\partial g_{0m}}{\partial x^n} \, \hat{P}^{mn}} \nonumber\\
   &-& \underline{\frac{1}{2} \left( \frac{\partial g_{00}}{\partial x^n}
      + \frac{1}{2} \frac{\partial g_{mm}}{\partial x^n} \right) \hat{P}^{0n}}
   \label{Hamsourcenl} \\
   &-& \nonumber
   \underline{\bigg[ \frac{1}{4} \frac{\partial (g_{\dObar{m}}-g_{\dUbar{m}})}{\partial x^m}
   + b_{\kappa m} \frac{\partial {b^\kappa}_n}{\partial x_n} \bigg] \hat{P}^{m0}} \\ \nonumber
   &+& \frac{1}{2} \left( b_{\kappa \Obar{n}} \frac{\partial {b^\kappa}_{\Ubar{n}}}{\partial x_n}
   - b_{\kappa \Ubar{n}} \frac{\partial {b^\kappa}_{\Obar{n}}}{\partial x_n} \right)
   \hat{J}^{(\Obar{n}\Ubar{n})n} \bigg\} \, d^3 x .
\end{eqnarray}
This contribution affects only the evolution equations (\ref{phatevol2}) for the conjugate momenta. For the linearized theory, these evolution equations simplify to
\begin{equation}\label{phatevollin}
   \frac{\partial \hat{p}^{\mu\nu}}{\partial x^0} =
   - \bar{b}^{\mu\kappa} \frac{\delta}{\delta {b^\kappa}_\nu}
      \left( \frac{H_{\rm source}}{\hbar c} + \left. \int X_{\alpha\beta}^B \, \hat{p}^{\alpha\beta}
      \, d^3 x \right|_{\hat{p}^{\alpha\beta}} \right) .
\end{equation}

The following identity is useful for carrying out the required functional derivatives:
\begin{eqnarray}
   \bar{b}^{\mu\lambda} \frac{\delta}{\delta{b^\lambda}_\nu}
   \int {b^\kappa}_\alpha \frac{\partial b_{\kappa \beta}}{\partial x^n} \, f \, d^3 x
   &=& - {\delta_\alpha}^\mu \, {\delta_\beta}^\nu \, \frac{\partial f}{\partial x^n} \nonumber\\
   && \hspace{-10em} + \;
   \mbox{$\bar{b}^\mu$}_{\kappa} \left( \frac{\partial {b^\kappa}_\beta}{\partial x^n}
   {\delta_\alpha}^\nu 
   - \frac{\partial {b^\kappa}_\alpha}{\partial x^n}
   {\delta_\beta}^\nu \right) f .
\label{auxderb}
\end{eqnarray}
where the function $f$ does not depend on the tetrad variables. By symmetrization in $\alpha$ and $\beta$ we obtain the corollary
\begin{equation}\label{auxderg}
   \bar{b}^{\mu\lambda} \frac{\delta}{\delta{b^\lambda}_\nu}
   \int \frac{\partial g_{\alpha\beta}}{\partial x^n} \, f \, d^3 x
   = - ({\delta_\alpha}^\mu \, {\delta_\beta}^\nu + {\delta_\alpha}^\nu \, {\delta_\beta}^\mu)
   \frac{\partial f}{\partial x^n} .
\end{equation}
For the linearized theory, only the term involving the derivative of $f$ in Eq.~(\ref{auxderb}) is relevant. We are now in a position to verify the consistency of the Hamiltonian formulation with Eqs.~(\ref{P0nevol})--(\ref{Pn0evol}).

The underlined terms in Eq.~(\ref{Hamsourcenl}) indicate contributions that are canceled by corresponding terms resulting from $X_{\alpha\beta}^B$ in the evolution equations. This cancellation is essential for ensuring that the Hamiltonian system correctly reflects the physical interpretation of the variables $\hat{p}^{\mu\nu}$.

The uncanceled  first four terms in the Hamiltonian (\ref{Hamsourcenl}) reproduce Eq.~(\ref{Pn0evol}). The evolution equation for $\hat{p}^{0n}$ becomes
\begin{equation}\label{p0nevol}
   \frac{\partial \hat{p}^{0n}}{\partial x^0} =
   - \frac{\partial \hat{p}^{mn}}{\partial x^m}
   + \frac{\partial \hat{p}^{00}}{\partial x_n}
   + \frac{\partial (\hat{p}^{nm}-\hat{P}^{nm})}{\partial x_m} .
\end{equation}
The last term vanishes by virtue of Eq.~(\ref{pPid}). While such cancellations play an essential role in relating the conjugate momenta to the external sources, we will often suppress them in subsequent expressions. Imposing the constraint $\hat{p}^{00}=0$, which is compatible with the evolution equation,
\begin{equation}\label{p00evol}
   \frac{\partial \hat{p}^{00}}{\partial x^0} =
   \frac{\partial (\hat{p}^{0n}-\hat{P}^{0n})}{\partial x^n} = 0 ,
\end{equation}
we recover the evolution equation (\ref{P0nevol}).

For the Hamiltonian evolution equation governing $\hat{p}^{mn}$, we find
\begin{eqnarray}
   \frac{\partial \hat{p}^{mn}}{\partial x^0} &=& \frac{1}{2} \bigg[
   \delta_{m\Ubar{n}} \, \frac{\partial \hat{J}^{(mn)\Obar{n}}}{\partial x^{\Obar{n}}}
   + \delta_{m\Obar{n}} \, \frac{\partial \hat{J}^{(mn)\Ubar{n}}}{\partial x^{\Ubar{n}}} \nonumber \\
   &-& \frac{\partial \hat{p}^{0n}}{\partial x^m}
   + (\delta_{m\Ubar{n}}-\delta_{m\Obar{n}}) \, \frac{\partial \hat{p}^{n0}}{\partial x^m} \bigg] ,
\label{pmnevol}
\end{eqnarray}
which implies Eq.~(\ref{Pmnasymevol}). It also implies the evolution equations for $\hat{p}^{\Obar{n}\Ubar{n}}+\hat{p}^{\Ubar{n}\Obar{n}}$, which we could not construct from the properties of the currents in Sec.~\ref{subsecJ}.

\subsection{Counting degrees of freedom}
The configurational degrees of freedom in the Hamiltonian formulation of composite gravity consist of the gauge vector fields $A_{a \mu}$ ($24$ in total) and the tetrad variables ${b^\kappa}_\mu$ ($16$ in total). These $40$ configurational variables are matched by the $40$ conjugate momenta $E^{a \mu}$ and ${p_\kappa}^\mu$, so that the total number of degrees of freedom is $80$. A substantial number of constraints is therefore required in order to obtain a physically viable theory of gravity.

As discussed in Sec.~\ref{secgaugefix}, we impose six primary gauge conditions on the tetrad variables. Their time derivatives must vanish, which yields six secondary constraints. We attempt to ensure that the Lorenz gauge conditions (\ref{Lorenzgauge}) arise as tertiary constraints. The quaternary constraints (\ref{gaugeconstrEa02}) follow from the field equations, so that we end up with a total of $18$ gauge constraints. The conditions (\ref{pPid}), together with the constraint $\hat{p}^{00}=0$, fix the $16$ conjugate momenta of the tetrad variables, thereby reducing the number of degrees of freedom from $80$ to $46$. With the six additional constraints required to ensure the proper evolution of $\hat{p}^{\Obar{n}\Ubar{n}}+\hat{p}^{\Ubar{n}\Obar{n}}$, we are left with $40$ degrees of freedom.

In Sec.~\ref{subsecexploicomp}, we identified $12$ primary constraints arising from the composition rule. To derive the higher‑order constraints implied by requiring their validity at all times, we have collected the relevant time derivatives in App.~\ref{apptimeder}.

Using the evolution equations (\ref{Atildeevol}) and (\ref{bbcorevol}), together with (\ref{Gammatildef}) and (\ref{bderivative}), we obtain the desired $12$ secondary constraints
\begin{equation}\label{secconstra}
   \tilde{E}_{(mn)j} = \frac{\partial \tilde{\Gamma}_{j0m}}{\partial x^n}
   - \frac{\partial \tilde{\Gamma}_{j0n}}{\partial x^m}
   + \Gamma^\sigma_{jm} \tilde{\Gamma}_{\sigma 0n} - \Gamma^\sigma_{jn} \tilde{\Gamma}_{\sigma 0m} ,
\end{equation}
and
\begin{equation}\label{secconstrb}
   \tilde{E}_{(0m)n} = \tilde{E}_{(0n)m} .
\end{equation}
It is quite remarkable that the Riemannian curvature tensor (\ref{Ftilform1}) appears in Eq.~(\ref{secconstra}) as an object defined in the background Minkowski space (see App.~\ref{appgeovars} for further discussion). According to Eq.~(\ref{Gammatildef}), each $\tilde{\Gamma}_{\sigma\mu n}$ can be expressed in terms of the configurational fields and their spatial derivatives. Consequently, the same is true for components $F_{amn}$ and $\tilde{F}_{\mu\nu mn}$ of the field tensor. The components $\tilde{F}_{0m0n}$, however, inevitably contain time derivatives of the configurational fields. In the Hamiltonian framework, these components are accessible as the conjugate momenta $\tilde{E}_{(0m)n}$.

Taking the time derivative of the secondary constraints (\ref{secconstra}), we use Eq.~(\ref{Etildeevol}) for the left-hand side. On the right-hand side, only time derivatives of the configurational variables appear, which can be eliminated by the respective evolution equations. We therefore write the tertiary constraints as
\begin{eqnarray}\label{terconstra}
   \frac{\partial \tilde{F}_{(mn)0j}}{\partial x_0} + \frac{\partial \tilde{F}_{(mn)ij}}{\partial x_i}
   &=& \\
   && \hspace{-6.5em} \eta^{\alpha\beta} \left( \Gamma^\sigma_{\alpha m} \tilde{F}_{(\sigma n)\beta j}
   + \Gamma^\sigma_{\alpha v} \tilde{F}_{(m \sigma)\beta j} \right) - \tilde{J}_{(mn)j} . \nonumber
\end{eqnarray}
These constraints correspond to nine of the Yang-Mills field equations given in Eq.~(\ref{METYMfieldeqs}). As discussed in Sec.~\ref{subsecJ}, ensuring the consistency of these equations with time evolution requires the conservation laws for the external currents. No further constraints arise for the configurational variables or for their conjugate momenta.

The time derivative of the secondary constraints (\ref{secconstrb}) can be evaluated by means of Eq.~(\ref{Etildeevol}),
\begin{eqnarray}
  \frac{\partial \tilde{F}_{(0m) jn}}{\partial x_j}
  &-& \eta^{\alpha\beta} \left( \Gamma^\sigma_{\alpha 0} \tilde{F}_{(\sigma m)\beta n}
  + \Gamma^\sigma_{\alpha m} \tilde{F}_{(0\sigma)\beta n} \right) + \tilde{J}_{(0m)n} \nonumber \\
  &=& \boxed{m \leftrightarrow n} \; .
\label{terconstrb}
\end{eqnarray}
By adding\begin{equation}\label{terconstrbsuppl}
   \frac{\partial \tilde{F}_{(0m)0n}}{\partial x_0} = \frac{\partial \tilde{F}_{(0n)0m}}{\partial x_0} ,
\end{equation}
to this equation, we recognize that the tertiary constraints (\ref{terconstrb}) are satisfied if three more Yang-Mills field equations hold. Again, no further constraints arise. In short, the $12$ tertiary constraints resulting from the composition rule express precisely the validity of the Yang-Mills field equations.

With a total of $36$ constraints imposed by the composition rule, only four degrees of freedom remain in the composite theory of gravity. This result, which is consistent with the analysis of the linearized theory of composite gravity with harmonic coordinate conditions \cite{hco240}, corresponds to two independent variables that satisfy second-order evolution equations.

In the case of planar wave propagation, these two variables in the solution (\ref{Amatrix}) are $h_{11}=-h_{22}$ and $h_{12}$. The isotropic static field surrounding a particle at rest is characterized by the three variables $a$, $b$, and $q$ in Eq.~(\ref{isoxb}). Using prior knowledge of the prefactor in Eq.~(\ref{Aisotropic}), the variables $a$, $b$, and $q$ can be determined from first-order differential equations.

\subsection{Comments on quantization}
The quantization of Yang-Mills-type theories is well-established for the electroweak and strong interactions \cite{PeskinSchroeder,WeinbergQFT2}. A robust approach to quantum field theory is provided by the philosophically founded particle picture of dissipative quantum field theory \cite{hcoqft,hco243}.

Since our Hamiltonian formulation of composite gravity is constructed in the combined space of tetrad and gauge vector fields, we expect two types of fundamental particles associated with gravitational interactions. The gauge vector fields $A_{a \mu}$ correspond to vector bosons of a Yang-Mills-type gauge theory, in analogy with photons, W and Z bosons, and gluons. It is natural to refer to these vector bosons as \textit{gravitons}, with the important clarification that they carry spin $1$. The tetrad variables ${b^\kappa}_\mu$ may be interpreted as double vector fields: they carry a vector index $\mu$ in the background Minkowski space and a vector index $\kappa$ in a local freely falling frame, which serves as an internal space associated with the local Lorentz symmetry group. We refer to the corresponding vector bosons as \textit{fallies} (singular \textit{fally}, pronounced with a short o as in \textit{folly}).

It may be surprising that two types of particles---gravitons and fallies---are needed to describe gravitational interactions. It should be emphasized, however, that composite gravity avoids the need for additional ghost particles that violate the spin–statistics theorem, which are commonly introduced in the quantization of Yang-Mills theories to handle gauge constraints within the BRST formalism (the acronym BRST refers to the work of Becchi, Rouet, Stora \cite{BecchiRouetStora76}, and independently Tyutin \cite{Tyutin75}). In the composite theory of gravity, the treatment of algebraic gauge constraints at the level of tetrad variables is significantly more direct and far simpler than in conventional gauge theories. Moreover, the tetrad fields offer the advantage of having a clear physical interpretation: they describe the transformation between freely falling local frames and the background Minkowski frame. Expressing also the gauge-vector fields for electroweak and strong interactions in terms of more fundamental fields may provide a promising path toward simpler gauge constraints and the avoidance of ghost particles in the Standard Model.

For the further discussion of the quantized composite theory of gravity, we adopt an approach that focuses on collisions between fundamental quantum particles \cite{hcoqft,hco243}. The essential features are the Fock space of free particles and stochastic jump processes that model collisions among them. According to the Yang-Mills contribution (\ref{Hfield}) to the Hamiltonian, gravitons participate in the usual three-particle interactions characteristic of gauge theories with non-commutative symmetry group. The tetrad contribution (\ref{Hamtetrad}) introduces collisions among fallies as well as interactions between fallies and gravitons. Finally, the source contribution (\ref{Hamsourceform}) couples the energy-momentum flux densities of other fundamental particles---such as leptons and quarks---to gravitons and fallies. Any appearance of the metric in the Hamiltonian signals an interaction involving a pair of fallies. Emission or absorption of a single fally is not possible.

According to the Englert-Brout-Higgs-Guralnik-Hagen-Kibble mechanism (or, more briefly, Higgs mechanism) for mass generation (see Secs.~20.1 and 20.2 of \cite{PeskinSchroeder} or Sec.~21.3 of \cite{WeinbergQFT2}), the masses of leptons and quarks are not intrinsic properties of these particles but instead arise from their interactions with the vacuum expectation value of the Higgs field. However, even in the absence of such a mechanism, a massless particle such as the photon can participate in gravitational interactions because it possesses a nonzero energy-momentum tensor.

Using a point-particle representation of quantum field theory has important advantages. Understanding the interaction of a gravitational field with a single point particle is sufficient to analyze its collisions with gravitons and fallies, and thus the interaction between gravitational fields and discrete matter distributions. Moreover, self-interaction effects can be handled more straightforwardly in a particle-based description than in a field-theoretic one. Previous work on composite gravity has discussed the conditions under which the motion of a point particle in a gravitational field is described by geodesic motion (see Sec. IV.G of \cite{hco240}).

\section{Conclusions and discussion}\label{secconclusions}
The freedom to choose among equivalent freely falling frames is the origin of the Lorentz group emerging as the local symmetry group in a theory of gravity. The freely falling frames are described by the tetrad variables, from which the gauge vector fields of a Yang-Mills-type theory can be constructed. Completing the theory requires the imposition of coordinate conditions, for which we have introduced the novel proposal (\ref{cocof}). For the gauge-fixing conditions, which can be imposed as algebraic constraints on the level of the tetrad variables, we propose a balance between generality and practicality.

The formulation of a gauge gravitation theory \cite{CapozzielloDeLau11,IvanenkoSar83} is a subtle matter. Immediately after the original work of Yang and Mills, Utiyama \cite{Utiyama56} considered the Yang-Mills theory based on the Lorentz group as a potential theory of gravity. His approach was criticized as ``unnatural'' by Yang (see footnote~5 of \cite{Yang74}). Yang's own allegedly more natural proposal \cite{Yang74} has, in turn, been criticized massively in Chapter~19 of \cite{BlagojevicHehl}. In this work, we achieve an appealing realization of the basic objective pursued by both Utiyama and Yang by consistently exploiting the concept of freely falling frames.

After presenting the complete set of field equations, we studied planar wave propagation and the static isotropic field around a point mass at rest. As in the case of electromagnetic waves, two transverse modes characterize the propagation of gravitational waves. Unlike the Schwarzschild solution of general relativity, our exact solution for the static isotropic field exhibits no singularities away from the location of the central mass.

In his pursuit of a unified field theory of gravity and electrodynamics, Einstein \cite{Einstein25} employs a variational principle on the combined spaces of metrics and connections, which he treats as independent variables. Within this approach, he allows not only for connections with torsion but also for an antisymmetric addition to the metric. He introduces an action that, in the linearized theory, leads to a Levi-Civita-type relation (\ref{Gammatildef}) between these two quantities. Within his ``true solution to the problem,'' the symmetric and antisymmetric parts of the geometric structures are then associated with gravitation and electrodynamics, respectively.

Similarly, we here use the combined space of tetrad and gauge-vector fields to develop a Hamiltonian formulation of a Yang-Mills-type theory of gravitation. In doing so, we convert a higher-derivative theory into a system of first-order differential equations.  A large number of constraints follow from the composition rule, leaving only four physical degrees of freedom in composite gravity. The Hamiltonian formulation presented in this paper is also essential for the quantization of the composite theory of gravity.

Returning to the Einstein quote given in the introduction---that only the sum (G) + (P) of a discretionarily chosen geometry (G) and appropriately adapted physical laws (P) is subject to experience---a delicate situation can arise. If one chooses a highly advanced geometry for a particular theory, say of gravity, and then attempts to unify it with another theory, such as that of electroweak interactions, which is typically and successfully formulated on a less sophisticated geometry (G'), this may prove to be impossible. The reason would be that an overly advanced geometry cannot be compensated for or ``undone'' by any physical laws. I am convinced that the most appropriate choice as a common ground for all known interactions is the Minkowski geometry (G'), as elaborated in this work.

\begin{acknowledgments}
I am grateful to the Department of Materials at ETH Zürich for continued support after my retirement.
\end{acknowledgments}

\appendix

\section{Structure of the Lorentz group}\label{AppstructureLG}
The structure constants of the Lorentz group can be expressed in the following explicit form (see Table~\ref{tabindexmatch} for the index conventions),
\begin{eqnarray}
   f^{abc} &=& \eta^{\kappa_a \lambda_c} \eta^{\kappa_b \lambda_a} \eta^{\kappa_c \lambda_b}
   - \eta^{\kappa_a \lambda_b} \eta^{\kappa_b \lambda_c} \eta^{\kappa_c \lambda_a}
   \nonumber\\
   &+& \eta^{\kappa_a \kappa_b} \big( \eta^{\kappa_c \lambda_a} \eta^{\lambda_b \lambda_c}
   - \eta^{\kappa_c \lambda_b} \eta^{\lambda_a \lambda_c} \big)
   \nonumber\\
   &+& \eta^{\kappa_a \kappa_c} \big( \eta^{\kappa_b \lambda_c} \eta^{\lambda_a \lambda_b}
   - \eta^{\kappa_b \lambda_a} \eta^{\lambda_b \lambda_c} \big)
   \nonumber\\
   &+& \eta^{\kappa_b \kappa_c} \big( \eta^{\kappa_a \lambda_b} \eta^{\lambda_a \lambda_c}
   - \eta^{\kappa_a \lambda_c} \eta^{\lambda_a \lambda_b} \big) .
\label{Lorentzstructure}
\end{eqnarray}
With this explicit representation, the following identity is obtained by straightforward summation,
\begin{equation}\label{supauxf1}
   f_{(\kappa\lambda)}^{bc} B_b C_c =
   \eta^{\kappa'\lambda'} \Big[ B_{(\kappa'\lambda)} C_{(\kappa\lambda')}
   - C_{(\kappa'\lambda)} B_{(\kappa\lambda')} \Big] .
\end{equation}
We then obtain
\begin{equation}\label{supauxf2}
   {b^\kappa}_\mu {b^\lambda}_\nu f_{(\kappa\lambda)}^{bc} B_b C_c =
   \bar{g}^{\rho\sigma} \Big[ \tilde{B}_{(\rho\mu)} \tilde{C}_{(\sigma\nu)}
   + \tilde{B}_{(\rho\nu)} \tilde{C}_{(\mu\sigma)} \Big] .
\end{equation}
This remarkably simple identity follows from the form of the structure constants of the Lorentz group.

\section{Geometric variables}\label{appgeovars}
Quite remarkable is the following deep relation between covariant derivatives associated with the Yang-Mills theory based on the Lorentz group on the one hand and covariant derivatives associated with connections on the other hand:
\begin{eqnarray}
   {b^\kappa}_\mu {b^\lambda}_\nu \left[
      \frac{\partial B_{(\kappa\lambda)}}{\partial x^\rho}
      + f_{(\kappa\lambda)}^{bc} A_{b\rho} B_c \right] &=& \nonumber\\
   && \hspace{-9em} \frac{\partial \tilde{B}_{(\mu\nu)}}{\partial x^\rho}
      - \Gamma^\sigma_{\rho\mu} \tilde{B}_{(\sigma\nu)}
      - \Gamma^\sigma_{\rho\nu} \tilde{B}_{(\mu\sigma)} , \qquad 
\label{central}
\end{eqnarray}
for an arbitrary quantity $B_a$ and the associated quantity
\begin{equation}\label{Btildef}
   \tilde{B}_{(\mu\nu)} = {b^\kappa}_\mu {b^\lambda}_\nu \, B_{(\kappa\lambda)} .
\end{equation}
The identity (\ref{central}) can be proven by simplifying its left-hand side by means of Eq.~(\ref{supauxf2}) and its right-hand side by means of Eq.~(\ref{Gammatildef}) and (\ref{Gammadef}).

By applying the relation (\ref{central}) to the definition (\ref{Fdefinition}) of the field tensor and repeatedly using Eq.~(\ref{Gammatildef}) we obtain the alternative representation
\begin{equation}\label{Ftilform1}
   \tilde{F}_{(\mu\nu)\alpha\beta} = R_{\mu\nu\alpha\beta} = \frac{\partial \tilde{\Gamma}_{\mu\nu\alpha}}{\partial x^\beta}
   - \frac{\partial \tilde{\Gamma}_{\mu\nu\beta}}{\partial x^\alpha}
   + \Gamma^\sigma_{\mu\alpha} \tilde{\Gamma}_{\sigma\nu\beta}
   - \Gamma^\sigma_{\mu\beta} \tilde{\Gamma}_{\sigma\nu\alpha} .
\end{equation}
This result shows that the transformed field tensor of the Yang-Mills theory is equal to the Riemann curvature tensor. As the latter tensor is usually written in terms of derivatives of the Christoffel symbol (see, e.g.\ \cite{Weinberg} or \cite{Jimenezetal19}), we use Eq.~(\ref{gderivative}) to obtain the alternative representation
\begin{equation}\label{Ftilform2}
   \bar{g}^{\mu\rho} R_{\rho\nu\alpha\beta} = 
   \frac{\partial \Gamma^\mu_{\nu\alpha}}{\partial x^\beta}
   - \frac{\partial \Gamma^\mu_{\nu\beta}}{\partial x^\alpha}
   + \Gamma^\mu_{\sigma\beta} \Gamma^\sigma_{\nu\alpha}
   - \Gamma^\mu_{\sigma\alpha} \Gamma^\sigma_{\nu\beta} .
\end{equation}

The Riemann curvature tensor (\ref{Ftilform2}) possesses the well-known symmetries 
\begin{equation}\label{Riemannsym}
   R_{\mu\nu\alpha\beta} = - R_{\nu\mu\alpha\beta} = - R_{\mu\nu\beta\alpha} = R_{\alpha\beta\mu\nu} ,
\end{equation}
and the cyclicity property
\begin{equation}\label{Riemanncyc}
   R_{\mu\nu\alpha\beta} + R_{\mu\beta\nu\alpha} + R_{\mu\alpha\beta\nu} = 0 ,
\end{equation}
(see, for example, Eqs.~(6.6.3)-(6.6.5) of \cite{Weinberg}). Note that Eq.~(\ref{Eamdef}) implies
\begin{equation}\label{EinFtilde}
   \tilde{E}_{(\mu\nu) j} = R_{\mu\nu j0} = - R_{\mu\nu 0j} = - R_{0j\mu\nu} ,
\end{equation}
so that Eq.~(\ref{Riemanncyc}) for $\mu=0$ becomes
\begin{equation}\label{Ecyc}
   \tilde{E}_{(mn)j} + \tilde{E}_{(nj)m} + \tilde{E}_{(jm)n} = 0 .
\end{equation}

The identity  (\ref{central}) also implies the transformed Yang-Mills field equations (\ref{YMfieldeqs}),
\begin{equation}\label{METYMfieldeqs}
   \frac{\partial \tilde{F}_{(\mu\nu)\sigma\rho}}{\partial x_\sigma} - \eta^{\alpha\beta} \left[ 
   \Gamma^\sigma_{\alpha\mu} \tilde{F}_{(\sigma\nu)\beta\rho}
   + \Gamma^\sigma_{\alpha\nu} \tilde{F}_{(\mu\sigma)\beta\rho} \right] =
   - \tilde{J}_{(\mu\nu)\rho} .
\end{equation}

Equations (\ref{central}) and (\ref{Ftilform2}) are quite remarkable. They suggest an intimate relationship between the Yang-Mills-type gauge theory with Lorentz symmetry group and Riemannian geometry. However, one should realize that the gauge theory of gravity nevertheless is fundamentally different from general relativity.
\begin{enumerate}
\item The gauge theory is defined on a background Minkowski space. Even though its field tensor is given by the Riemann curvature tensor, the metric does not affect the underlying flat spacetime. In particular, all volume integrals are performed in the flat spacetime.
\item In systems consisting of point particles, discrete sums occur instead of the integrals associated with field theories, thus avoiding the issue of metric effects on integrals. This observation might be relevant to fundamental particle physics.
\item The field equations (\ref{METYMfieldeqs}) of the gauge theory involves derivatives of the field tensor on the left-hand side and the currents (\ref{JhatfromT}), which contain derivatives of the energy-momentum-flux tensor, on the right-hand side. In general relativity, the field equation is an algebraic relation between the curvature tensor and the energy-momentum-flux tensor.
\item The origin of the occurrence of geometric variables is Einsteins equivalence principle. The idea of freely falling frames is implemented in the Yang-Mills-type theory with Lorentz symmetry group.
\end{enumerate}

\section{Perturbation theory}\label{secperturb}
We now construct a perturbation expansion in terms of the energy-momentum flux tensor ${T_\mu}^\nu$ or, more conveniently, in terms of
\begin{equation}\label{Sdef}
   {S_\mu}^\nu = {T_\mu}^\nu - \frac{1}{2} {\delta_\mu}^\nu \,  {T_\rho}^\rho .
\end{equation}
Note the identity ${S_\rho}^\rho = -{T_\rho}^\rho$.

By means of Eq.~(\ref{Ftilform1}), the field equations (\ref{METYMfieldeqs}) can be rewritten as
\begin{eqnarray}
   && \hspace{-2em}
   \frac{\partial^2 \tilde{\Gamma}_{\alpha\beta\nu}}{\partial x^\mu \partial x_\alpha} -
   \frac{\partial^2 \tilde{\Gamma}_{\alpha\beta\mu}}{\partial x^\nu \partial x_\alpha} =
   \frac{\partial}{\partial x_\alpha} \left( 
   \Gamma^\sigma_{\alpha\mu} \tilde{\Gamma}_{\sigma\nu\beta}
   - \Gamma^\sigma_{\alpha\nu} \tilde{\Gamma}_{\sigma\mu\beta} \right) \qquad \nonumber \\
   &-& \eta^{\alpha\alpha'} \left( \Gamma^\sigma_{\alpha\mu} \tilde{F}_{(\sigma\nu)\alpha'\beta}
   - \Gamma^\sigma_{\alpha\nu} \tilde{F}_{(\sigma\mu)\alpha'\beta} \right) + \tilde{J}_{(\mu\nu)\beta} .
\label{METYMfieldeqsalt}
\end{eqnarray}

By means of Eq.~(\ref{JhatfromT}) for the conserved currents, the first-order equations can be written in the form of integrability conditions,
\begin{equation}\label{fieldeq1}
   \frac{\partial}{\partial x^\mu} \left(
   \frac{\partial \tilde{\Gamma}_{\alpha\beta\nu}}{\partial x_\alpha}
   - \frac{8 \pi G}{c^4} S_{\nu\beta} \right) =
    \frac{\partial}{\partial x^\nu} \left(
   \frac{\partial \tilde{\Gamma}_{\alpha\beta\mu}}{\partial x_\alpha}
   - \frac{8 \pi G}{c^4} S_{\mu\beta} \right) .
\end{equation}
They imply the existence of potentials $\phi_\beta$ with
\begin{equation}\label{fieldeq1int}
   \frac{\partial \tilde{\Gamma}_{\alpha\beta\mu}}{\partial x_\alpha}
   - \frac{8 \pi G}{c^4} S_{\mu\beta} = \frac{\partial\phi_\beta}{\partial x^\mu} 
   = \frac{\partial^2 \phi}{\partial x^\beta \partial x^\mu} ,
\end{equation}
where the last equality follows from the symmetry of the left-hand side in $\beta$ and $\mu$. This integration brings the field equations to the same order of derivatives as those of general relativity. Indeed, for $\phi = {g_\alpha}^\alpha/2$, we obtain
\begin{equation}\label{fieldeq1intGR}
   \frac{\partial^2 g_{\beta\mu}}{\partial x_\alpha \partial x^\alpha}
   - \frac{\partial^2 g_{\alpha\beta}}{\partial x_\alpha \partial x^\mu}
   - \frac{\partial^2 g_{\alpha\mu}}{\partial x_\alpha \partial x^\beta}
   + \frac{\partial^2 {g_\alpha}^\alpha}{\partial x^\beta \partial x^\mu}
   = - \frac{16 \pi G}{c^4} S_{\mu\beta} ,
\end{equation}
which is the linearized version of Einstein's field equation. In view of the successful predictions of general relativity for the gravitational far-field properties, this is an expected result.

The left-hand side of Eq.~(\ref{METYMfieldeqsalt}) is cylic in $\mu$, $\nu$, $\beta$. To first order, the right-hand side shares this property, but at higher orders it does not. Although cyclicity of the right-hand side is not manifest in the form of the field equations, any admissible solution must nevertheless respect it.

Once $\tilde{\Gamma}_{\alpha\beta\nu}$ and $g_{\alpha\beta}$ have been determined to order $N$, the right-hand side of Eq.~(\ref{METYMfieldeqsalt}) can be evaluated to order $N+1$. This allows us to determine the perturbative solution to order $N+1$.

\section{Some useful time derivatives}\label{apptimeder}
The evolution equations
\begin{eqnarray}\label{Atildeevol}
   \frac{\partial \tilde{A}_{(\mu\nu)n}}{\partial x^0} &=& \Gamma^\sigma_{0\mu} \tilde{A}_{(\sigma\nu)n}
   + \Gamma^\sigma_{0\nu} \tilde{A}_{(\mu\sigma)n}
   - \tilde{E}_{(\mu\nu)n} \\
   && \hspace{-3em} + \; \frac{\partial \tilde{A}_{(\mu\nu)0}}{\partial x^n}
   - \bar{b}^\sigma{}_\kappa \frac{\partial {b^\kappa}_\mu}{\partial x^n} \tilde{A}_{(\sigma\nu)0}
   + \bar{b}^\sigma{}_\kappa \frac{\partial {b^\kappa}_\nu}{\partial x^n} \tilde{A}_{(\sigma\mu)0} ,
   \nonumber
\end{eqnarray}
follow from Eqs.~(\ref{Aevolaj}) and (\ref{central}). From the definition of $ X_{\mu\nu}$ in Eq.~(\ref{Xmunu}) we obtain
\begin{equation}\label{bbcorevol}
   \frac{\partial}{\partial x^0}
   \left( b_{\kappa\mu} \frac{\partial {b^\kappa}_\nu}{\partial x^n} \right) =
   \frac{\partial X_{\mu\nu}}{\partial x^n}
   - \bar{b}^\sigma{}_\kappa \frac{\partial {b^\kappa}_\mu}{\partial x^n} X_{\sigma\nu}
   + \bar{b}^\sigma{}_\kappa \frac{\partial {b^\kappa}_\nu}{\partial x^n} X_{\sigma\mu} .
\end{equation}
These two evolution equations can be combined into
\begin{equation}\label{Gamtilevol}
   \frac{\partial \tilde{\Gamma}_{\mu\nu n}}{\partial x^0} = \tilde{E}_{(\mu\nu) n}
   + \frac{\partial \tilde{\Gamma}_{\mu\nu 0}}{\partial x^n}
   - \Gamma^\sigma_{\nu 0} \tilde{\Gamma}_{\sigma\mu n}
   + \Gamma^\sigma_{\mu 0} \tilde{\Gamma}_{\sigma\nu n} .
\end{equation}

For the transformed evolution equations (\ref{Eevolaj}) we obtain
\begin{eqnarray}\label{Etildeevol}
   \frac{\partial \tilde{E}_{(\mu\nu)n}}{\partial x^0} &=&
   - \frac{\partial \tilde{F}_{(\mu\nu) jn}}{\partial x_j} \\
   &+& \eta^{\alpha\beta} \left( \Gamma^\sigma_{\alpha\mu} \tilde{F}_{(\sigma\nu)\beta n}
   + \Gamma^\sigma_{\alpha\nu} \tilde{F}_{(\mu\sigma)\beta n} \right) \nonumber
   - \tilde{J}_{(\mu\nu)n} .
\end{eqnarray}

% \bibliography{hcopubs}


\end{document}